\documentclass{bmcart}

\usepackage{amsthm,amsmath}
\usepackage{amssymb}
\usepackage{dsfont}
\usepackage[utf8]{inputenc}
\usepackage{url}
\usepackage{color}
\usepackage{xspace}
\usepackage{graphicx}
\usepackage{hyperref}
\usepackage{musicography}
\usepackage{pdfpages}
\usepackage{relsize}
\usepackage{siunitx}

\graphicspath{{figures/}{figures/violin_scattering/}{figures/flute_scattering/}{./}}
\startlocaldefs

\DeclareMathOperator*{\median}{median}
\DeclareMathOperator*{\card}{card}

\newcommand*{\eg}{e.g.,\@\xspace}
\newcommand*{\ie}{i.e.,\@\xspace}

\newcommand{\lnameref}[1]{%
\bgroup
\let\nmu\MakeLowercase
\nameref{#1}\egroup}
\newcommand{\fnameref}[1]{%
\bgroup
\def\nmu{\let\nmu\MakeLowercase}%
\nameref{#1}\egroup}

\newcommand{\nmu}{}

\endlocaldefs

\begin{document}

\begin{frontmatter}

\begin{fmbox}
\dochead{Research}


\title{Time--Frequency Scattering Accurately Models Auditory Similarities Between Instrumental Playing Techniques}


\author[
   addressref={aff2},                   
   email={vincent.lostanlen@ls2n.fr}            
]{\inits{VL}\fnm{Vincent} \snm{Lostanlen}}
\author[
   addressref={aff2},
   email={christian.elhajj@ls2n.fr}
]{\inits{CE}\fnm{Christian} \snm{El-Hajj}}
\author[
   addressref={aff3},
   email={mathias.rossignol@lonofi.com}
]{\inits{MR}\fnm{Mathias} \snm{Rossignol}}
\author[
   addressref={aff3},
   email={gregoire.lafay@lonofi.com}
]{\inits{GL}\fnm{Gr\'egoire} \snm{Lafay}}
\author[
   addressref={kth,ccm},
   email={janden@kth.se}
]{\inits{JA}\fnm{Joakim} \snm{And\'en}}
\author[
  corref={aff2},
  addressref={aff2},
  email={mathieu.lagrange@ls2n.fr}
]{\inits{ML}\fnm{Mathieu} \snm{Lagrange}}


\address[id=aff2]{%
  \orgname{LS2N, CNRS, Centrale Nantes, Nantes University},
  \street{1, rue de la Noe},
  \postcode{44000},
  \city{Nantes},
  \cny{France}
}
\address[id=aff3]{%
  \orgname{Lonofi},
  \street{57 rue Letort},
  \postcode{75018},
  \city{Paris},
  \cny{France}
}
\address[id=kth]{%
  \orgname{Department of Mathematics, KTH Royal Institute of Technology}
  \street{Lindstedtsv\"{a}gen 25}
  \city{SE-100 44, Stockholm}
  \cny{Sweden}
}
\address[id=ccm]{%
  \orgname{Center for Computational Mathematics, Flatiron Institute},
  \street{162 5th Avenue},
  \city{New York, NY 10010},
  \cny{USA}
}


\begin{artnotes}
\end{artnotes}

\end{fmbox}


\begin{abstractbox}

\begin{abstract}
Instrumental playing techniques such as vibratos, glissandos, and trills often denote musical expressivity, both in classical and folk contexts.
However, most existing approaches to music similarity retrieval fail to describe timbre beyond the so-called ``ordinary'' technique, use instrument identity as a proxy for timbre quality, and do not allow for customization to the perceptual idiosyncrasies of a new subject.
In this article, we ask 31 human participants to organize 78 isolated notes into a set of timbre clusters.
Analyzing their responses suggests that timbre perception operates within a more flexible taxonomy than those provided by instruments or playing techniques alone.
In addition, we propose a machine listening model to recover the cluster graph of auditory similarities across instruments, mutes, and techniques.
Our model relies on joint time--frequency scattering features to extract spectrotemporal modulations as acoustic features.
Furthermore, it minimizes triplet loss in the cluster graph by means of the large-margin nearest neighbor (LMNN) metric learning algorithm. 
Over a dataset of 9346 isolated notes, we report a state-of-the-art average precision at rank five (AP@5) of $\mathsf{99.0\%\pm1}$.
An ablation study demonstrates that removing either the joint time--frequency scattering transform or the metric learning algorithm noticeably degrades performance.
\end{abstract}


\end{abstractbox}

\end{frontmatter}


\section*{Introduction}
\label{sec:intro}

Music information retrieval (MIR) operates at two levels: symbolic and auditory \cite{downie2003mir}.
By relying on a notation system, the symbolic level allows the comparison of musical notes in terms of quantitative attributes, such as duration, pitch, and intensity at the source.
Timbre, in contrast, is a qualitative attribute of music, and is thus not reducible to a one-dimensional axis \cite{siedenburg2019chapter}.
As a result, symbolic representations describe timbre indirectly, either via visuotactile metaphors (\eg{} bright, rough, and so forth \cite{faure1996icmpc}) or via an instrumental playing technique (\eg{} bowed or plucked) \cite{lostanlen2018extended}.

Despite their widespread use, purely linguistic references to timbre fail to convey the intention of the composer.
On the one hand, adjectives such as \emph{bright} or \emph{rough} are prone to misunderstanding, as they do not prescribe any musical gesture that is capable of achieving them \cite{antoine2018isma}.
On the other hand, the sole mention of a playing technique does not specify its effect in terms of auditory perception.
For instance, although the term \emph{breathy} alludes to a playing technique that is specific to wind instruments, a cellist may accomplish a seemingly breathy timbre by bowing near the fingerboard, i.e., \emph{sul tasto} in the classical terminology.
Yet, in a diverse instrumentarium, the semantic similarity between playing technique denominations does not reflect such acoustical similarity \cite{kolozali2011ismir}.

Although a notation-based study of playing techniques in music has research potential in music information retrieval \cite{calvo2020acm}, the prospect of modeling timbre perception necessarily exceeds the symbolic domain.
Instead, it involves a cognitive process which arises from the subjective experience of listening \cite{erickson1975book}.
The simulation of this cognitive process amounts to the design of a multidimensional feature space wherein some distance function evaluates pairs of stimuli.
Rather than merely discriminating instruments as mutually exclusive categories, this function must reflect judgments of acoustic dissimilarity, all other parameters---duration, pitch, and intensity---being equal \cite{thoret2018jasa}.

\subsection*{Use case}

Behind the overarching challenge of coming up with a robust predictive model for listening behaviors in humans, the main practical application of timbre similarity retrieval lies in the emerging topic of computer-assisted orchestration \cite{maresz2013cmr}.
In such context, the composer queries the software with an arbitrary audio signal.
The outcome is another audio signal which is selected from a database of instrumental samples and perceptually similar to the query.

The advantage of this search is that, unlike the query, the retrieved sound is precisely encoded in terms of duration, pitch, intensity, instrument, and playing technique.
Thus, following the aesthetic tradition of spectralism in contemporary music creation, the computer serves as a bridge from the auditory level to the symbolic level, \ie{} from a potentially infinite realm of timbral sensations to a musical score of predefined range \cite{caetano2019swarm}.

Composers may personalize their search engine by manually defining a cluster graph via the Cyberlioz interface (see \lnameref{sec:data-collection} section).
This preliminary annotation lasts $30$ to $60$ minutes, which is relatively short in comparison with the duration of the $N^\prime = 9346$ audio samples in the database: i.e., roughly three hours of audio.

\subsection*{Goal}
This article proposes a machine listening system which computes the dissimilarity in timbre between two audio samples.\footnote{
For the sake of research reproducibility, the source code for the experimental protocol of this paper is available online, alongside anonymized data from human subjects:  \url{https://github.com/mathieulagrange/lostanlen2020jasmp}}
Crucially, this dissimilarity is not evaluated in terms of acoustic tags, but in terms of \emph{ad hoc} clusters, as defined by a human consensus of auditory judgments.
Our system consists of two stages: unsupervised feature extraction and supervised metric learning.
The feature extraction stage is a nonlinear map which relies on the joint time--frequency scattering transform \cite{anden2015mlsp,anden2019tsp}, followed by per-feature Gaussianization \cite{lostanlen2018jasmp}.
It encodes patterns of spectrotemporal modulation in the acoustic query while offering numerical guarantees of stability to local deformations \cite{anden2012scattering}.
The metric learning stage is a linear map, optimized via large-margin nearest neighbors (LMNN) \cite{weinberger2009distance}.
It reweights scattering coefficients so that pairwise distances between samples more accurately reflect human judgments on a training set.
These human judgments may be sourced from a single subject or the intersubjective consensus of multiple participants \cite{mcadams1995psychres}.

Figure \ref{fig:pipeline} summarizes our experimental protocol: it illustrates how visual annotations (top) can inform feature extraction (center) to produce a nearest-neighbor search engine which is consistent with human judgments of timbre similarity (bottom).

\subsection*{Approach}
The main contribution of this article can be formulated as the intersection between three topics.
To the best of our knowledge, prior literature has addressed these topics separately, but never in combination.

First, our dataset encompasses a broad range of extended playing techniques, well beyond the so-called ``ordinary'' mode of acoustic production.
Specifically, we fit pairwise judgments for 78 different techniques arising from 16 instruments, some of which include removable timbre-altering devices such as mutes.

Secondly, we purposefully disregard the playing technique metadata underlying each audio sample during the training phase of our model.
In other words, we rely on listeners, not performers, to define and evaluate the task at hand.

Thirdly, we supplement our quantitative benchmark with visualizations of time--frequency scattering coefficients in the rate--scale domain for various typical samples of instrumental playing techniques.
These visualizations are in line with visualizations of the modulation power spectrum in auditory neurophysiology \cite{patil2012ploscompbiol}, while offering an accelerated algorithm for scalable feature extraction.

Our paper strives to fill the gap in scholarship between MIR and music cognition in the context of extended playing techniques.
From the standpoint of MIR, the model presented here offers an efficient and generic multidimensional representation for timbre similarity, alongside theoretical guarantees of robustness to elastic deformations in the time--frequency domain.
Conversely, from the standpoint of music cognition, our model offers a scalable and biologically plausible surrogate for stimulus-based collection of acoustic dissimilarity judgments, which is readily tailored to subjective preferences.


\section*{\nmu Related work}
\label{sec:related-work}

Timbre involves multiple time scales in conjunction, from a few microseconds for an attack transient to several seconds for a sustained tone.
Therefore, computational models of timbre perception must summarize acoustic information over a long analysis window \cite{joder2009taslp}.
Mapping this input to a feature space in which distances denote timbral dissimilarity requires a data-driven stage of dimensionality reduction.
In this respect, the scientific literature exhibits a methodological divide as regards the collection of human-annotated data \cite{siedenburg2016jnmr}: while the field of MIR mostly encodes timbre under the form of ``audio tags'', music psychology mostly measures timbre similarity directly from pairwise similarity judgments.

\subsection*{Automatic classification of musical instruments and playing techniques}
On the one hand, most publications in music information retrieval cast timbre modeling as an audio classification problem  \cite{martin1998asa,brown1999jasa,eronen2000icassp,herrera2003jnmr,wieczorkowska2003jiis,livshin2004dafx,krishna2004icassp,kaminskyj2005jiis,benetos2006icassp,bhalke2016jiis}.
In this context, the instrumentation of each musical excerpt serves as an unstructured set of ``audio tags,'' encoded as binary outputs within some predefined label space.
Because such tags often belong to the metadata of music releases, the process of curating a training set for musical instrument classification requires little or no human intervention.
Although scraping user-generated content from online music platforms may not always reflect the true instrumentation with perfect accuracy, it offers a scalable and ecologically valid insight onto the acoustic underpinnings of musical timbre.

Furthermore, supplementing user-generated content with the outcome of a crowdsourced annotation campaign allows an explicit verification of instrument tags.
For instance, the Open-MIC dataset \cite{humphrey2018ismir}, maintained by the Community for Open and Sustainable Music Information Research (COSMIR) \cite{mcfee2016ismir}, comprises a vast corpus of 20k polyphonic music excerpts spanning 20 instruments as a derivative of the Free Music Archive (FMA) dataset \cite{defferrard2017ismir}. 
Another example is the Medley-solos-DB dataset \cite{lostanlen2016ismir}, which comprises 21k monophonic excerpts from eight instruments as a derivative of the MedleyDB dataset of multitrack music \cite{bittner2014ismir}.

Over the past decade, the availability of large digital audio collections, together with the democratization of high-performance computing on dedicated hardware, has spurred the development of deep learning architectures in music instrument recognition \cite{mcfee2015ismir,pons2017eusipco,gururani2018ismir}.
Notwithstanding the growing accuracy of these architectures in the large-scale data regime, it remains unclear how to extend them from musical instrument recognition to playing technique recognition, where labeled samples are considerably more scarce \cite{loureiro2004ismir}.
We refer to \cite{han2017taslp} for a recent review of the state of the art in this domain.

\subsection*{Spectrotemporal receptive fields (STRF) in music cognition}
On the other hand, the field of music psychology investigates timbre with the aim of discovering its physiological and behavioral foundations \cite{mcadams2009chapter}.
In this setting, prior knowledge of instrumentation, however accurate, does not suffice to conduct a study on timbre perception: rather, the timbre perception relies on an interplay of acoustic and categorical information \cite{siedenburg2016frontiers}.
Yet, collecting subjective responses to acoustic stimuli is a tedious and unscalable procedure, which restricts the size of the musical corpus under study.
These small corpus sizes hamper the applicability of optimization algorithms for representation learning, such as stochastic gradient descent in deep neural networks.

While training artificial neurons is prone to statistical overfitting, advanced methods in electrophysiology allow to observe the firing patterns of biological neurons in the presence of controlled stimuli.
This observation, originally carried out on the ferret, has led to a comprehensive mapping of the primary auditory cortex in terms of its spectrotemporal receptive fields (STRFs) \cite{depireux2001jneur}.
The STRF of a neuron is a function of time and frequency which represents the optimal predictor of its post-stimulus time histogram during exposure to a diverse range of auditory stimuli \cite{aertsen1981biolcyb}.
The simplest method to compute it in practice is by reverse correlation, i.e.\ by averaging all stimuli that trigger an action potential \cite{deboer1968biomed}.
Historically, STRFs were defined by their Wigner--Ville distribution \cite{flandrin1998book}, thereby sparing the choice of a tradeoff in time--frequency localization, but eliciting cross-term interferences \cite{eggermont1993hearing}.
Since then, the STRF of a neuron was redefined as a spectrographic representation of its spike-triggered average \cite{klein2000compneur}.

Although this new definition is necessarily tied to a choice of spectrogram parameters, it yields more interpretable patterns than a Wigner--Ville distribution.
In particular, a substantial portion of spectrographic STRFs exhibit a ripple-like response around a given region $(t, \lambda)$ of the time--frequency domain \cite{theunissen2000jneur}.
This response can be approximately described by a pair of scalar values: a temporal modulation rate $\alpha$ in Hertz and a frequential modulation rate $\beta$ in cycles per octave.

Interestingly, both $\alpha$ and $\beta$ appear to be arranged in a geometric series and independent from the center time $t$ and center frequency $\lambda$.
This observation has led auditory neuroscientists to formulate an idealized computational model for STRF, known as the ``full cortical model'' \cite{chi2005jasa}, which densely covers the rate--scale domain $(\alpha, \beta)$ using geometric progressions.
Because they do not require a data-driven training procedure, STRF yield a useful form of domain-specific knowledge for downstream machine listening applications, especially when the number of annotated samples is relatively small.

\subsection*{Spectrotemporal receptive fields (STRFs) as a feature extractor}
Over recent years, several publications have employed the full cortical model as a feature extractor for a task of musical instrument classification, both in isolated recordings \cite{patil2012ploscompbiol} and in solo phrases \cite{patil2015eurasip}.
These biologically inspired features outperform the state of the art, especially in the small data regime where deep learning is inapplicable.
Furthermore, the confusion matrix of the full cortical model in the label space of musical instruments is strongly correlated with the confusion matrix between a human listener and the ground truth.
Another appeal of the full cortical model is that the three-way tensor of frequency $\lambda$, rate $\alpha$, and scale $\beta$ can be segmented into contiguous regions of maximal perceptual relevance for each instrument \cite{thoret2016jasa}.
This is unlike fully end-to-end learning architectures, whose post hoc interpretability requires advanced techniques for feature inversion \cite{mishra2018ismir}.
Lastly, beyond the realm of supervised classification, a previous publication \cite{hemery2015frontiers} has shown that query-by-example search with STRFs allows to discriminate categories of environmental soundscapes, even after temporal integration and unsupervised dimensionality reduction.

The reasons above make STRFs an appealing feature extractor for a perceptual description of timbral similarity across instrumental playing techniques.
Nonetheless, current implementations of STRF suffer from a lack of scalability, which explains why they have found few applications in MIR thus far.
Indeed, the full cortical model is usually computed via two-dimensional Fourier transforms over adjacent time--frequency regions, followed by averaging around specific rates and scales.
This approach requires a uniform discretization of the scalogram, and thus an oversampling of the lower-frequency subbands to the Nyquist frequency of the original signal.
In contrast, joint time--frequency scattering offers a faster extraction of spectrotemporal modulations while preserving properties of differentiability \cite{andreux2020jmlr} and invertibility \cite{lostanlen2019dafx}.
Such acceleration is made possible by discretizing the wavelet transforms involved in time--frequency scattering according to a multirate scheme, both along the time and the log-frequency variables \cite{anden2019tsp}.
In this multirate scheme, every subband is discretized at its critical sample rate, i.e., in proportion to its center frequency.
As a by-product, the multirate approach draws an explicit connection between scattering networks and deep convolutional networks, because both involves numerous convolutions with small kernels, pointwise rectifying nonlinearities, and pooling operations \cite{mallat2016philtrans}.

Moreover, a quantitative benchmark over Medley-solos-DB has demonstrated that joint time--frequency scattering, unlike purely temporal scattering, outperforms deep convolutional networks in supervised musical instrument classification, even in a relatively large data regime with 500 to 5k samples per class \cite{anden2019tsp}. 
However, it remains to be seen whether joint time--frequency scattering is capable of fine-grained auditory categorization, involving variability in instrument, mute, and playing technique.
In addition, previous publications on joint time--frequency scattering lack a human-centric evaluation, independently from any classification task.
Beyond the case of STRF, we refer to \cite{caetano2019chapter} for a detailed review of the state of the art on audio descriptors of timbre.

\section*{\nmu Perceptual data collection}
\label{sec:data-collection}

The philharmonic orchestra encompasses four families of instruments: strings, woodwinds, brass, and percussion.
In this article, we focus on the first three, and leave the question of learning auditory similarities between percussion instruments to future research.
We refer to \cite{wu2018taslp} and \cite{pearce2019appliedsciences} for reviews of the recent literature on the timbre modeling of percussive instruments, from the standpoints of MIR and music cognition, respectively.

\subsection*{Dataset}
We consider a list of 16 instruments: violin (Vn), viola (Va), cello (Vc), contrabass (Cb), concert harp (Hp), Spanish guitar (Gtr), accordion (Acc), flute (Fl), soprano clarinet (BbCl), alto saxophone (ASax), oboe (Ob), bassoon (Bn), trumpet in C (TpC), French horn (Hn), tenor trombone (TTbn), and bass tuba (BBTb).
Among this list, the first six are strings, the next six are woodwind, and the last four are brass.
Some of these instruments may be temporarily equipped with timbre-altering mutes, such as a rubber sordina on the bridge of a violin or an aluminium ``wah-wah'', also known as \emph{harmon}, inside in the bell of a trumpet.
Once augmented with mutes, the list of 16 instruments grows to 33.
Furthermore, every instrument, whether equipped with a mute or not, affords a panel of playing techniques ranging in size between 11 (for the accordion) and 41 (for the bass tuba).
In the rest of this paper, we abbreviate instrument--mute--technique by means of the acronym ``IMT''.
One example of IMT is \texttt{TpC+S-ord}, \ie{} trumpet in C with a straight mute played in the ordinary technique.
Another example of IMT is \texttt{Vn-pont}, \ie{} violin without any mute played in the \emph{sul ponticello} technique (bowing near the bridge).

Performers can play each IMT at various pitches according to the tessitura of their instrument.
This tessitura may depend on the choice of playing technique but is independent of the choice of mute.
Among the 16 instruments in this study, the two instruments with widest and narrowest tessituras, in their respective ordinary techniques, are the accordion (81 semitones) and the trumpet in C (32 semitones) respectively.
Lastly, each IMT may be played at up to five intensity dynamics, ranging from quietest to loudest as: pianissimo (\emph{pp}), piano (\emph{p}), mezzo forte (\emph{mf}), forte (\emph{f}), and fortissimo (\emph{ff}).
The resort to a non-ordinary playing technique may restrict both the tessitura and the dynamics range of the instrument--mute pair under consideration.
For example, the pitch of pedal tones in brass instruments is tied to the fundamental mode of the bore, \ie{} usually $\mathsf{B}\musFlat$ or $\mathsf{F}$.
Likewise, the intensity of key clicks in the oboe is necessarily \emph{pp}, while the intensity of snap pizzicato \emph{\`a la Bart\'ok} in plucked strings is necessarily \emph{ff}.

In summary, audio signals from isolated musical notes may vary across three categorical variables (instrument, mute, and technique) and two quantitative variables (intensity and pitch).
The Studio On Line dataset (SOL), recorded at Ircam in 1998, offers a joint sampling of these variables.
The version of SOL that we use throughout this paper, named ``0.9 HQ'', amounts to a total of 25444 audio signals.
Beyond playing techniques, we should note that SOL erases other factors of acoustic variability, such as identity of performer, identity of instrument manufacturer, audio acquisition equipment, and room response characteristics, which are all restricted to singletons.
Addressing these factors of variability is beyond the scope of this paper, which focuses on the influence of playing technique.
Despite this restriction, the SOL dataset remains impractically large for collecting human similarity judgments.
Our protocol addresses this problem by means of three complementary approaches: disentanglement of factors, expert pre-screening, and the use of an efficient annotation interface.

\subsection*{Disentanglement of factors}
First, we purposefully disentangle categorical variables (IMTs) from continuous variables (pitch and intensity) in the SOL dataset.
Indeed, under first approximation, the perception of timbre is invariant to pitch and intensity.
Therefore, we select auditory stimuli according to a reference pitch and a reference intensity; in our case, middle C ($\mathrm{C_4}$) and \emph{mf}.
After this selection, every IMT triplet contains a single acoustic exemplar, regarded as canonical in the following.
The number of canonical stimuli for the entire SOL dataset is equal to 235.
We should note, however, that the proposed pitch and intensity cannot be strictly enforced across all IMTs.
Indeed, as explained above, a fraction of IMTs can only be achieved at restricted values of pitch and intensity parameters, \eg{} pedal tones or key clicks.
Therefore, at a small cost of consistency, we only enforce the pitch--intensity reference (\ie{} $\mathrm{C_4}$ and \emph{mf}) when practically feasible, and fall back to other pitches and intensities if necessary.

\subsection*{Expert pre-screening}
Secondly, we reduce the number of IMTs in our study by focusing on those which are deemed to be most relevant.
Here, we define the relevance of an IMT as the possibility of imitating it by means of another IMT from a different instrument.
One example of such imitation is the acoustic similarity between slap tonguing in reed instruments and a snap pizzicato in string instruments.
To collect perceptual ratings of relevance, we recruited two professors in music composition at the Paris Conservatory (CNSMDP\footnote{CNSMDP: Conservatoire National Sup\'erieur de Musique et de Danse de Paris. Official website: \url{https://www.conservatoiredeparis.fr}}).
Each of them inspected the entire corpus of $235$ IMTs and annotated them in terms of relevance according to a Likert scale with seven ticks.
In this Likert scale, the value 1 (least relevant) denotes that the IMT under consideration has a timbre that is idiosyncratic, and that therefore, it is unlikely that humans will pair it with other IMTs.
Conversely, the value 7 (most relevant) denotes that the IMT under consideration bears a strong similarity with some other IMT from the corpus.

Once both experts completed their annotations, we retained all IMTs whose average score was judged equal to 3 or higher, thus resulting in a shortlist of $N=78$ IMTs (see Tables 1 and 2 in the Appendix).
It is worth noting that, according to both experts, the timbre of the accordion was judged too idiosyncratic to be relevant for this experiment, regardless of playing technique.
Indeed, the accordion is the only instrument in the aforementioned list of
instrument to have free reeds, keyboard-based actuation, or handheld airflow.
Consequently, regardless of mute and technique, the set of instruments $\mathcal{I}$ in our study contains 15 elements.

\subsection*{Efficient annotation interface}
Thirdly, we design a graphical user interface for partitioning a corpus of short audio samples.
The need for such an interface arises from the unscalability of Likert scales in the context of pairwise similarity judgments.
Assuming that similarity is a symmetric quantity, collecting a dense matrix of continuously valued ratings of similarity among a dataset of $N$ items would require $\frac{1}{2}(N^2-N)$ Likert scales.
In the case of $N=78$ IMTs, the task would amount to about 3k horizontal sliders, \ie{} several hours of cumbersome work for the human annotator. 

Engaging as many participants as possible in our study called for a more streamlined form of human--computer interaction, even if it sacrificed the availability of continuously valued ratings.
To this end, we implemented a web application, named Cyberlioz, in which the user can spontaneously listen and arrange sounds into clusters of timbre similarity.\footnote{The Web application for efficient audio annotation, as well as the raw anonymized responses of all 31 participants to our study, is available at: \url{https://soundthings.org/research/cyberlioz/}}
The name Cyberlioz is a portmanteau between the prefix cyber- and the French composer Hector Berlioz.
The choice is by no means coincidental: Berlioz is famous for having, in his \emph{Treatise on Orchestration} (1844), shed a particular focus on the role of timbre as a parameter for musical expression.

Cyberlioz consists of a square panel on which is displayed a collection of circular grey dots, each of them corresponding to one of the IMTs, and initially distributed uniformly at random.
Hovering the screen pointer onto each dot results in a playback of a representative audio sample of this IMT, \ie{} $\mathrm{C}_4$ and \emph{mf} in most cases. Furthermore, each dot can be freely placed on the screen by clicking, dragging, and dropping.
Lastly, the user can assign a color to each dot among a palette of 20 hues.
The goal of the Cyberlioz interface is to form clusters of timbre similarity between IMTs, expressed by sameness of color.

Cyberlioz implements a data collection procedure known as ``free sorting''. In comparison with the direct collection of timbre dissimilarity ratings, free sorting is more efficient yet less accurate \cite{giordano2011multivariate}. We refer to \cite{elliott2013jasa} for an example protocol in which timbre similarity judgments rely on stimuli pairs rather than on a free sorting task.

In comparison with web-based forms, Cyberlioz offers a more intuitive and playful user experience, while limiting the acquisition of similarity judgments to a moderate duration of $30$ to $60$ minutes for each participant.
Another advantage of Cyberlioz is that it allows to present all stimuli at once rather than according to a randomized sequence.

In May and June 2016, we recruited volunteers to use Cyberlioz on their own computers, via a web browser, and equipped with a pair of earphones. The subjects were asked to ‘cluster sounds into groups by assigning the same color to the corresponding dots according to how similar the sounds are’.

We publicized this study on the internal mailing list of students at CNSMDP, as well as two international mailing lists for research in music audio processing: AUDITORY  and ISMIR Community.\footnote{
For more information about these mailing lists, please visit:
\url{http://www.auditory.org/} and \url{http://ismir.net/}}
Within two months, $K=31$ participants accessed Cyberlioz and completed the task.

Personal information on the age, sex, nor musical background of participants is not collected, because the goal of our perceptual study is to build a consensus of similarity judgments, rather than to compare demographic subgroups.

In particular, we leave the important question of the effect of musical training on the perception of auditory similarities between playing techniques as future work.

\subsection*{Hypergraph partitioning}

Once the data collection campaign was complete, we analyzed the color assignments of each subject $k$ and converted them into a cluster graph $\mathcal{G}_k$, where the integer $k$ is an anonymized subject index, ranging between $1$ and $K$.
For a given $k$, the graph $\mathcal{G}_k$ contains $N$ vertices, each representing a different IMT in the corpus.
In $\mathcal{G}_k$, an edge connects any two vertices $m$ and $n$ if the corresponding dots in Cyberlioz have the same color.
Otherwise, there is no edge connecting $m$ and $n$.
Thus, $\mathcal{G}_k$ contains as many connected components as the number of similarity clusters for the subject $k$, \ie{} the number of distinct colors on the Cyberlioz interface in the response of $k$.

For a particular subject $k$, let us denote by $C_k$ the number of clusters in the graph $\mathcal{G}_k$.
Figure \ref{fig:cluster-histograms}(a) shows the histogram of $C_k$ across the cohort of $K=31$ participants.
We observe that the number of clusters varies between $3$ and $19$ with a median value of $10$.
Accordingly, the number of samples belonging to a cluster varies between $1$ (the most frequent value) and $50$, as shown in Figure \ref{fig:cluster-histograms}(b).

We aggregate the similarity judgments from all $K$ participants by embedding them into a hypergraph $\mathcal{H}$, that is, a graph whose edges may connect three or more vertices at once.
Specifically, $\mathcal{H}$ contains $N$ vertices, each representing an IMT; and each ``hyperedge'' in $\mathcal{H}$ corresponds to some connected component in one of the graphs $\mathcal{G}_1, \ldots, \mathcal{G}_K$.
Then, we convert the hypergraph $\mathcal{H}$ back into a conventional graph $\mathcal{G}_0$ by means of a combinatorial optimization algorithm known as hypergraph partitioning \cite{kernighan1970efficient}.

To construct $\mathcal{G}_0$, we select a number of clusters that is equal to the maximal value of the $C_k$'s; that is, $C_0 = 19$.
Then, we run hypergraph partitioning on $\mathcal{H}$ to assign each vertex $i$ to one of the $C_0$ clusters in $\mathcal{G}_0$.
Intuitively, hypergraph partitioning optimizes a tradeoff between two objectives: first, balancing the size of all clusters in terms of their respective numbers of vertices; and secondly, keeping most hyperedges enclosed within as few distinct clusters as possible \cite{han1997scalable,strehl2002cluster}.

While the graphs $\mathcal{G}_1, \ldots, \mathcal{G}_K$ encode the subjective similarity judgments of participants $1$ to $K$, the graph $\mathcal{G}_0$ represents a form of consensual judgment that is shared across all participants while discarding intersubjective variability.
Although the rest of our paper focuses on the consensus $\mathcal{G}_0$, it is worth pointing out that the same technical framework could apply to a single subject $k$, or to a subgroup of the $K=31$ participants.
This remark emphasizes the potential of our similarity learning method as a customizable tool for visualizing and extrapolating the timbre similarity space of a new subject.

\section*{\nmu Machine listening methods}
\label{sec:methods}
The previous section described our protocol for collecting timbral similarity judgments between instrumental playing techniques.
In this section, we aim to recover these similarity judgments from digital audio recordings according to a paradigm of supervised metric learning.
To this end, we present a machine listening system composing joint time--frequency scattering and LMNN.

\subsection*{Joint time--frequency scattering transform}
\label{sec:scattering}

Let $\boldsymbol{\psi} \in \mathbf{L}^2(\mathbb{R}, \mathbb{C})$ be a complex-valued filter with zero average, dimensionless center frequency equal to one, and an equivalent rectangular bandwidth (ERB) equal to $1/Q$.
We define a constant-$Q$ wavelet filterbank as the family $\boldsymbol{\psi}_{\lambda} : t \mapsto \lambda \boldsymbol{\psi}(\lambda t)$.
Each wavelet $\boldsymbol{\psi}_{\lambda}$ has a center frequency of $\lambda$, an ERB of $\lambda/Q$, and an effective receptive field of $(2\pi Q/\lambda)$ in the time domain.
In practice, we define $\boldsymbol{\psi}$ as a Morlet wavelet:
\begin{equation}
\boldsymbol{\psi}:t \longmapsto \exp\left(-\dfrac{t^2}{2\sigma_{\psi}^2}\right)
\big( \exp\left(2\pi \mathrm{i}t\right) - \kappa_{\psi} \big),
\label{eq:psi}
\end{equation}
where the Gaussian width $\sigma_{\psi}$ grows in proportion with the quality factor $Q$ and the corrective term $\kappa_{\psi}$ ensures that $\boldsymbol{\psi}$ has a zero average.
Moreover, we discretize the frequency variable $\lambda$ according to a geometric progression of common ratio $2^{\frac{1}{Q}}$.
Thus, the base-two logarithm of center frequency, denoted by $\log_2 \lambda$, follows an arithmetic progression.
We set the constant quality factor of the wavelet filterbank $(\boldsymbol{\psi}_\lambda)_\lambda$ to $Q=12$, thus matching twelve-tone equal temperament in music.

Convolving the wavelets in this filterbank with an input waveform $\boldsymbol{x}\in\mathbf{L}^2(\mathbb{R})$, followed by an application of the pointwise complex modulus yields the wavelet scalogram
\begin{equation}
\mathbf{U_1}\boldsymbol{x}(t,\lambda) =
\big\vert
\boldsymbol{x}
\ast
\boldsymbol{\psi}_{\lambda}
\big\vert(t)
=
\left\vert
\int_{\mathbb{R}}
\boldsymbol{x}(t - t^{\prime})
\,
\boldsymbol{\psi}_{\lambda} (t^{\prime})
\;
\mathrm{d}{t^\prime}
\right\vert,
\label{eq:U1}
\end{equation}
which is discretized similarly to the constant-$Q$ transform of \cite{schorkhuber2010smc}.
Then, we define a two-dimensional Morlet wavelet $\boldsymbol{\Psi}$ of the form
\begin{equation}
\Psi : (t, u) \longmapsto
\exp\left(-\dfrac{t^2+u^2}{2\sigma_{\Psi}^2}\right)
\Big( \exp\big(2\pi \mathrm{i} (t + u)\big) - \kappa_{\Psi} \Big),
\label{eq:Psi}
\end{equation}
taking two real variables $t$ and $u$ as input.
In the rest of this paper, we shall refer to $\boldsymbol{\Psi}$ as a time--frequency wavelet.
The former is the time variable while the latter is the base-two logarithm of frequency: $u = \log_2 \lambda$.
Note that $u$ roughly corresponds to the human perception of relative pitch \cite{lostanlen2020icassp}.

We choose the Gaussian width $\sigma_{\Psi}$ in Equation \ref{eq:Psi} such that the quality factor of the Morlet wavelet $\Psi$ is equal to one, both over the time dimension and over the log-frequency dimension.
Furthermore, the corrective term $\kappa_{\Psi}$ ensures that $\boldsymbol{\Psi}$ has a zero average over $\mathbb{R}^2$, similarly to Equation \ref{eq:psi}.
From $\mathbf{\Psi}$, we define a two-dimensional wavelet filterbank of the form:
\begin{equation}
\mathbf{\Psi}_{\alpha,\beta} : (t, u) \longmapsto
\alpha \, \beta \, \mathbf{\Psi}(\alpha t, \beta u).
\end{equation}
In the equation above, $\alpha$ is a temporal modulation rate and $\beta$ is a frequential modulation scale, following the terminology of spectrotemporal receptive fields (STRF, see \lnameref{sec:related-work} section).
While $\alpha$ is measured in Hertz and is strictly positive, $\beta$ is measured in cycles per octaves and may take positive as well as negative values.
Both $\alpha$ and $\beta$ are discretized by geometric progressions of common ratio equal to two.
Furthermore, the edge case $\beta=0$ corresponds to $\mathbf{\Psi}_{\alpha,\beta}$ being a Gaussian low-pass filter over the log-frequency dimension, while remaining a Morlet band-pass filter of center frequency $\alpha$ over the time dimension.
We denote this low-pass filter by $\boldsymbol{\phi}_F$ and set its width to $F=2$ octaves.

We now convolve the scalogram $\mathbf{U_1}$ with time--frequency wavelets $\mathbf{\Psi}_{\alpha,\beta}$ and apply the complex modulus, yielding the four-way tensor
\begin{align}
\mathbf{U_2}\boldsymbol{x} (t, \lambda, \alpha, \beta)
=&
\big \vert
\mathbf{U_1}\boldsymbol{x} \circledast \boldsymbol{\Psi}_{\alpha,\beta}
\big \vert (t, \lambda)
\nonumber \\
=&
\left\vert
\iint_{\mathbb{R}^2}
\mathbf{U_1}\boldsymbol{x}(t - t^\prime, \log \lambda - u^\prime)\,
\boldsymbol{\Psi}_{\alpha,\beta}(t^\prime, u^\prime)\;
\mathrm{d}t^\prime \, \mathrm{d}u^\prime
\right\vert,
\label{eq:U2}
\end{align}
where the circled asterisk operator $\circledast$ denotes a joint convolution over time and log-frequency.
In the equation above, the sample rate of time $t$ is proportional to $\alpha$.
Conversely, the sample rate of log-frequency $u = \log_2 \lambda$ is proportional to $\vert\beta\vert$ if $\beta \neq 0$ and proportional to $F^{-1}$ otherwise.

Let $\boldsymbol{\phi}_T$ be a Gaussian low-pass filter.
We define the joint time--frequency scattering coefficients of the signal $\boldsymbol{x}$ as the four-way tensor
\begin{align}
\mathbf{S_2}(t, \lambda, \alpha, \beta)
=&
\mathbf{U_2}\boldsymbol{x}
\circledast
\big(\boldsymbol{\phi}_T  \otimes \boldsymbol{\phi}_F\big)
(t, \lambda)
\nonumber \\
=&
\iint_{\mathbb{R}^2}
\mathbf{U_1}\boldsymbol{x}(t - t^\prime, \log \lambda - u^\prime, \alpha, \beta)
\,\boldsymbol{\phi}_T (t^\prime)
\,\boldsymbol{\phi}_F (u^\prime)
\;\mathrm{d}t^\prime \, \mathrm{d}u^\prime,
\label{eq:S2}
\end{align}
where the symbol $\otimes$ denotes the outer product over time and log-frequency.
In the equation above, the sample rate of time $t$ is proportional to $T^{-1}$ and the sample rate of log-frequency $u=\log_2 \lambda$ is proportional to $F^{-1}$.
Furthermore, the rate $\alpha$ spans along a geometric progression ranging from $T^{-1}$ to $\lambda/Q$.
In the following, we set the time constant to $T=\SI{1000}{\milli\second}$ unless specified otherwise.

The tensor $\mathbf{S_2}$ bears a strong resemblance with the idealized response of an STRF at the rate $\alpha$ and the scale $\beta$.
Nevertheless, in comparison with the ``full cortical model'' \cite{patil2012ploscompbiol}, joint time--frequency scattering enjoys a thirty-fold reduction in dimensionality while covering a time span that is four times larger (1000 ms) and an acoustic bandwidth that is also four times larger (0--16 kHz).
This is due to the multirate discretization scheme applied throughout the application of wavelet convolutions and pointwise modulus nonlinearities.

In addition to second-order scattering coefficients (Equation \ref{eq:S2}), we compute joint time--frequency scattering at the first order by convolving the scalogram $\mathbf{U_1 \boldsymbol{x}}$ (Equation \ref{eq:U1}) with the low-pass filter $\boldsymbol{\phi}_T$ over the time dimension with wavelets $\boldsymbol{\psi}_{\beta}$ ($\beta \geq0$) over the log-frequency dimension, and by applying the complex modulus:
\begin{align}
\mathbf{S_1}\boldsymbol{x}(t, \lambda, \alpha=0, \beta) &=
\big\vert
\mathbf{U_1}\boldsymbol{x} \circledast
\big(\boldsymbol{\phi}_T \otimes \boldsymbol{\psi}_{\beta} \big)
\big\vert (t, \lambda)
\nonumber \\
&=
\left \vert
\iint_{\mathbb{R}^2}
\mathbf{U_1}\boldsymbol{x}(t - t^\prime, \log \lambda - u^\prime)
\,\boldsymbol{\phi}_T (t^\prime)
\,\boldsymbol{\psi}_{\beta} (u^\prime)
\;\mathrm{d}t^\prime \, \mathrm{d}u^\prime
\right \vert,
\label{eq:S1}
\end{align}
where the time constant $T$ is the same as in Equation $\ref{eq:S2}$, \ie{} $T=\SI{1000}{\milli\second}$ by default.

Over the time variable, we set the modulation rate $\alpha$ of $\mathbf{S_1}$ to zero in the equation above.
Conversely, over the log-frequency variable, the edge case $\beta=0$ corresponds to replacing the wavelet $\boldsymbol{\psi}_{\beta}$ by the low-pass filter $\boldsymbol{\phi}_F$.
We refer to \cite{anden2019tsp} for more details on the implementation of joint time--frequency scattering.

We adopt the multi-index notation $p = (\lambda, \alpha, \beta)$ as a shorthand for the tuple of frequency, rate, and scale.
The tuple $p$ is known as a scattering path (see \cite{mallat2012cpam}), and may apply to index both first-order ($\mathbf{S_1}$) and second-order ($\mathbf{S_2}$) coefficients.
Given an input waveform $\boldsymbol{x}$, we denote by $\mathbf{S}\boldsymbol{x}$ the feature vector resulting from the concatenation of $\mathbf{S_1}\boldsymbol{x}$ and $\mathbf{S_2}\boldsymbol{x}$:
\begin{equation}
\mathbf{S}\boldsymbol{x}\big(t, p=(\lambda, \alpha, \beta)\big)
=\left\{
\begin{array}{ll}
\mathbf{S_{1}}\boldsymbol{x}(t,\lambda,\alpha,\beta) & \textrm{if }\alpha=0,\\
\mathbf{S_{2}}\boldsymbol{x}(t,\lambda,\alpha,\beta) & \textrm{otherwise.}
\end{array}\right.
\end{equation}

\subsection*{Median-based logarithmic compression and affine standardization}
\label{sec:gaussianization}


Now, we apply a pointwise nonlinear transformation on averaged joint time-frequency scattering coefficients.
The role of this transformation, which is adapted to the dataset in an unsupervised way, is to Gaussianize the histogram of amplitudes of each scattering path $p$.
We consider a collection $\mathcal{X}$ of $N$ waveforms $\boldsymbol{x}_1,\ldots,\boldsymbol{x}_N$. For every path $p$ in the joint time--frequency scattering transform operator $\mathbf{S}$, we average the response of each scattering coefficient $\mathbf{S}\boldsymbol{x}_n$ over time and take its median value across all samples $n$ from $1$ to $N$:
\begin{equation}
\boldsymbol{\mu}(p) =
\median_{1\leq n \leq N}
\int_{\mathbb{R}} \mathbf{S}\boldsymbol{x}_n(t, p) \;\mathrm{d}t.
\end{equation}
If the collection is split between a training set and a test set (see \lnameref{sec:results} section), we compute $\boldsymbol{\mu}$ on the training set only.
Then, to match a decibel-like perception of loudness, we apply the following adaptive transformation, which composes a median-based renormalization and a logarithmic compression:
\begin{equation}
\mathbf{\widetilde{S}}\boldsymbol{x}_n (p) =
\log \left(
1 +
\dfrac{
\int_\mathbb{R} \mathbf{S}\boldsymbol{x}_n (t, p) \;\mathrm{d}t
}{\varepsilon\boldsymbol{\mu}(p)}
\right)
\label{eq:S_tilde}
\end{equation}
where $\varepsilon$ is a predefined constant.
The offset of one before the application of the pointwise logarithm ensures that the transformation is nonexpansive in the sense of Lipschitzian maps: there exists a constant $c$ such that
\begin{equation}
\big \Vert
\mathbf{\widetilde{S}}\boldsymbol{x}_m
-
\mathbf{\widetilde{S}}\boldsymbol{x}_{n}
\big \Vert
\leq
c
\big \Vert
\mathbf{S}\boldsymbol{x}_m
-
\mathbf{S}\boldsymbol{x}_{n}
\big \Vert
\end{equation}
for every pair of samples $(\boldsymbol{x}_m, \boldsymbol{x}_n)$.
On a dataset of environmental audio recordings, a previous publication has shown empirically that Equation \ref{eq:S_tilde} brings the histogram of $\mathbf{\widetilde{S}}\boldsymbol{x}_n (p)$ closer to a Gaussian distribution \cite{lostanlen2018jasmp}.
Since then, this finding has also been confirmed in the case of musical sounds \cite{lostanlen2018extended}.

Lastly, we standardize every feature $\mathbf{\widetilde{S}}\boldsymbol{x}_n$ to null mean and unit variance, across the dataset $\mathcal{X} = \{\boldsymbol{x}_1 \ldots \boldsymbol{x}_N\}$, independently for each scattering path $p$.
Again, if $\mathcal{X}$ is split between training and test sets, we measure means and variances over the training set only and propagate them as constants to the test set.
With a slight abuse of notation, we still denote by $\mathbf{\widetilde{S}}\boldsymbol{x}_n (p)$ the standardized log-scattering features at path $p$ for sample $n$, even though its value differs from Equation \ref{eq:S_tilde} by an affine transformation.

\subsection*{Metric learning with large-margin nearest neighbors (LMNN)}
Let $\boldsymbol{x}$ be some arbitrary audio sample in the dataset $\mathcal{X}$.
Let $\mathcal{G}$ be a cluster graph with $N = \card \mathcal{X}$ vertices and $C$ clusters in total.
We denote by $\mathcal{G}(\boldsymbol{x})$ the cluster to which the sample $\boldsymbol{x}$ belongs.
Given another sample $\boldsymbol{y}$ in $\mathcal{X}$, $\boldsymbol{y}$ is similar to $\boldsymbol{x}$ if and only if belongs to the cluster $\mathcal{G}(\boldsymbol{x})$.
Because $\mathcal{G}$ is a disjoint union of complete graphs, this relation is symmetric: $\boldsymbol{x} \in \mathcal{G}(\boldsymbol{y})$ is equivalent to $\boldsymbol{y} \in \mathcal{G}(\boldsymbol{x})$.

In our protocol, $\boldsymbol{x}$ contains the sound of an isolated musical note and the cluster graph $\mathcal{G}$ encodes auditory similarities within the dataset $\mathcal{X}$.
Moreover, we take $\mathcal{G}$ to be the equal to the ``consensus'' cluster graph $\mathcal{G}_0$, i.e., arising from the partition of a hypergraph $\mathcal{H}$ which contains the judgments of all $K$ participants from our perceptual study (see \lnameref{sec:data-collection} section).


We denote by $\mathbf{\widetilde{S}}\boldsymbol{x}$ the feature vector of joint time--frequency scattering resulting from $\boldsymbol{x}$.
This vector includes both first-order and second-order scattering coefficients, after median-based logarithmic compression and affine standardization (see subsections above).
Furthermore, we denote by $\mathcal{Y}_R (\boldsymbol{x})$ the list of $R$ nearest neighbors to $\mathbf{\widetilde{S}}\boldsymbol{x}$ in the feature space of joint time--frequency scattering coefficients according to the Euclidean metric.
Unlike cluster similarity, this relationship is not symmetric: $\boldsymbol{y}\in\mathcal{Y}_R (\boldsymbol{x})$ does not necessarily imply $\boldsymbol{x}\in\mathcal{Y}_R (\boldsymbol{y})$.
Note that the dependency of $\mathcal{Y}_R (\boldsymbol{x})$ upon the operator $\mathbf{S}$ is left implicit. 
In all of the following, we set the constant $R$ to $5$; this is in accordance with our chosen evaluation metric, average precision at rank 5 (AP@5, see \lnameref{sec:results} section).

Let $P$ be the number of scattering paths in the operator $\mathbf{S}$. The LMNN algorithm learns a matrix $\mathbf{L}$ with $P$ rows and $P$ columns by minimizing an error function of the form:
\begin{equation}
\mathcal{E}(\mathbf{L}) = \frac{1}{2} \mathcal{E}_{\textrm{pull}} (\mathbf{L}) + \frac{1}{2} \mathcal{E}_{\textrm{push}} (\mathbf{L})
\end{equation}
where, intuitively, $\mathcal{E}_{\textrm{pull}}$ tends to shrink local Euclidean neighborhoods in feature space while $\mathcal{E}_{\textrm{push}}$ tends to penalize small distances between samples that belong to different clusters in $\mathcal{G}$.

The definition of $\mathcal{E}_{\textrm{pull}}$ is:
\begin{equation}
\mathcal{E}_{\textrm{pull}} (\mathbf{L}) =
\sum_{\boldsymbol{x}\in\mathcal{X}}
\sum_{\boldsymbol{y}\in\mathcal{Y}_R (\boldsymbol{x})}
\big\Vert
\mathbf{L}\mathbf{\widetilde{S}}\boldsymbol{x} - \mathbf{L}\mathbf{\widetilde{S}}\boldsymbol{y}
\big\Vert^2,
\end{equation}
Note that the error term $\mathcal{E}_\textrm{pull}$ is unsupervised, in the sense that it does not depend on the cluster assignments of $\boldsymbol{x}$ and $\boldsymbol{y}$ in $\mathcal{G}$.

While the term $\mathcal{E}_{\textrm{pull}}$ operates on pairs of samples, the term $\mathcal{E}_{\textrm{push}}$, operates on triplets $(\boldsymbol{x}, \boldsymbol{y}, \boldsymbol{z})\in\mathcal{X}^3$.
The first sample, $\boldsymbol{x}$, is known as an ``anchor.''
The second sample, $\boldsymbol{y}$, is known as a ``positive'', and is assumed to belong to the Euclidean neighborhood of the anchor: $\boldsymbol{y} \in \mathcal{Y}_R (\boldsymbol{x})$.
The third sample, $\boldsymbol{z}$, is known as a ``negative'', and is assumed to belong to a different similarity cluster as the anchor: $\boldsymbol{z}\not\in\mathcal{G}(\boldsymbol{x})$.
The term $\mathcal{E}_{\textrm{push}}$ penalizes $\mathbf{L}$ unless the positive-to-anchor distance is smaller than the negative-to-anchor distance by a margin of at least $1$.

The definition of $\mathcal{E}_{\textrm{push}}$ is:
\begin{equation}
\mathcal{E}_{\textrm{push}} (\mathbf{L}) =
\sum_{\boldsymbol{x}\in\mathcal{X}}
\sum_{\boldsymbol{y}\in\mathcal{Y}_R (\boldsymbol{x})}
\sum_{\boldsymbol{z}\not\in\mathcal{G}(\boldsymbol{x})}
\rho
\Big(
1 +
\big\Vert
\mathbf{L}\mathbf{\widetilde{S}}\boldsymbol{x} - \mathbf{L}\mathbf{\widetilde{S}}\boldsymbol{y}
\big\Vert^2
-
\big\Vert
\mathbf{L}\mathbf{\widetilde{S}}\boldsymbol{x} - \mathbf{L}\mathbf{\widetilde{S}}\boldsymbol{z}
\big\Vert^2
\Big),
\end{equation}
where the function $\rho : u \mapsto \max (0, u)$ denotes the activation function of the rectified linear unit (ReLU), also known as hinge loss.
The cost function described in the equation above is known in deep learning as ``triplet loss'' and has recently been applied to train large-vocabulary audio classifiers in an unsupervised way \cite{jansen2018icassp}.
We refer to \cite{bellet2015book} for a review of the state of the art in metric learning.

\subsection*{Extension to diverse pitches and dynamics}

In order to suit the practical needs of contemporary music composers, computer-assisted orchestration must draw from a diverse realm of instruments and techniques.
Therefore, whereas our data collection procedure for timbre similarity judgments focused on a single pitch (middle C) and a single intensity level (\emph{mf}), we formulate our machine listening experiment on an expanded dataset of audio samples, containing variations in pitch and dynamics.

Given an audio stimulus $\boldsymbol{x}_n$ from our perceptual study, we seek its position in the cluster graph $\mathcal{G}_0$.
Then, we identify its IMT triplet, scrape for audio samples in SOL matching this triplet, and assign them all to the same cluster as the original audio stimulus.
We repeat the same procedure for all $N=78$ nodes in $\mathcal{G}_0$, resulting in $N^\prime = 9346$ samples in total.
Thus, from a limited amount of human annotation, we curate a relatively large subset of SOL, amounting to about one third of the entire dataset ($9346$ out of $25444$ samples).

In doing so, we assume that the perception of timbre is fully invariant to frequency transposition as well as changes in dynamics.
This assumption coincides with the commonplace definition of timbre as an autonomous parameter of musical sound.
Previous studies on real-world musical sounds have confirmed that listeners are able to ignore salient pitch differences while rating timbre similarity \cite{handel2001musicperception,marozeau2003dependency}, insofar as these differences do not exceed one octave.
Since then, another study has shown that trained musicians can identify the similarity within pairs of notes from two instruments (horn and bassoon) spanning a range of $2.5$ octaves with an accuracy of $80\%$ \cite{steele2006musicperception}.
In comparison, non-musicians has great difficulty identifying whether two notes were produced by the same instrument or not when the notes were separated by one octave.
Thus, the influence of pitch on timbre perception appears to depend upon musical training.
With this important caveat in mind, we leave as future work the question of disentangling the effects of pitch and musical training in the modeling of auditory similarities between instrumental playing techniques.

\subsection*{Evaluation metric}

Let us denote by $\boldsymbol{x}_1, \ldots, \boldsymbol{x}_{{N}^{\prime}}$ the $N^{\prime}$ audio samples associated with our annotated dataset.
Given a sample $n$ and a human subject $k$, we denote by $\mathcal{G}_k$ the cluster graph associated to the subject $k$, and by $\mathcal{G}_k (n)$ the cluster to which the sample $\boldsymbol{x}_{n}$ belongs.
Our machine listening system takes the waveform $\boldsymbol{x}_{n}$ as input and returns a ranked list of nearest neighbors: $\mathbf{\Phi}_1 (\boldsymbol{x}_n)$, $\mathbf{\Phi}_2 (\boldsymbol{x}_n)$, $\mathbf{\Phi}_3 (\boldsymbol{x}_n)$, and so forth.

In the context of browsing an audio collection by timbre similarity, $\boldsymbol{x}_n$ is a user-defined query while the function $\mathbf{\Phi}$ plays the role of a search engine.
We consider the first retrieved sample, $\mathbf{\Phi}_1 (\boldsymbol{x}_n)$, to be relevant to user $k$ if and only if it belongs to the same cluster as $\boldsymbol{x}_n$ in the cluster graph $\mathcal{G}_k$; hence the Boolean condition $\mathbf{\Phi}_{1}(\boldsymbol{x}_{n}) \in \mathcal{G}_k (n)$.
Likewise, the second retrieved sample is relevant if and only if $\mathbf{\Phi}_{2}(\boldsymbol{x}_{n}) \in \mathcal{G}_k (n)$.
To evaluate $\mathbf{\Phi}$ on the query $\boldsymbol{x}_n$, we measure the relevance of all nearest neighbors $\mathbf{\Phi}_r (\boldsymbol{x}_n)$ up to some fixed rank $R$ and average the result:
\begin{equation}
p_{\mathbf{\Phi}}(n, k, R) =
    \dfrac{1}{R}
    \sum_{r=1}^{R}
    \mathlarger{\mathds{1}}
    \big(
        \mathbf{\Phi}_r(\boldsymbol{x}_n)
        \in
        \mathcal{G}_k (n)
    \big).
\end{equation}
In the equation above, the indicator function $\mathds{1}$ converts Booleans to integers, i.e., $\mathds{1}(b)$ returns one if $b$ is true and return zero if $b$ is false.
Thus, the function $p_{\mathbf{\Phi}}$ takes fractional values between $0$ and $1$, which are typically expressed in percentage points.

The precision at rank $R$ of the system $\mathbf{\Phi}$ is defined as the average value taken by the function $p_{\mathbf{\Phi}}$ over the entire corpus of $N^{\prime}$ audio samples:
\begin{equation}
\mathrm{P}_{\mathbf{\Phi}}(k, R) =
\dfrac{1}{N^{\prime}}
\sum_{n=1}^{N^{\prime}}
p_{\mathbf{\Phi}}(n, k, R)
\end{equation}

Lastly, the ``average precision at rank $R$'' (henceforth, AP@$R$) is the average value of $\mathrm{P}_{\mathbf{\Phi}}$, for constant $R$, across all $K=31$ participants from our perceptual study:
\begin{equation}
\mathrm{AP}_{\mathbf{\Phi}}(R) =
\dfrac{1}{K}
\sum_{k=1}^{K}
\mathrm{P}_{\mathbf{\Phi}}(k, R)
\end{equation}
It appears from the above that an effective system $\mathbf{\Phi}$ should retrieve sounds whose IMT triplets are similar according to all of the $K$ cluster graphs $\mathcal{G}_1 \ldots \mathcal{G}_K$.

In the rest of this paper, we set $R$ to $5$.
This is in accordance with the protocol of \cite{lostanlen2018extended}, in which the authors trained a metric learning algorithm on the SOL dataset to search for similar instruments and playing techniques, yet without the intervention of a human subject.

\section*{\nmu Results}
\label{sec:results}

The previous section described our methods for extracting spectrotemporal modulations in audio signals, as well as learning a non-Euclidean similarity metric between them.
We now turn to apply these methods to the problem of allocating isolated musical notes to clusters of some timbre similarity graph $\mathcal{G}$.
In practice, for training purposes, the cluster graph $\mathcal{G}$ represents the consensus of the $K=31$ clustering provided by the users interacting with the Cyberlioz web application, which was described in the \lnameref{sec:data-collection} section ($\mathcal{G}=\mathcal{G}_0$).
However, for evaluation purposes, this cluster graph corresponds to the subjective preferences of a single user $k\geq 1$, in which case we take $\mathcal{G}=\mathcal{G}_k$.

\subsection*{Best performing system}
Our best performing system comprises five computational blocks:

\begin{enumerate}
\item Joint time--frequency scattering up to a maximal time scale of $T=$ \SI{1000}{\milli\second},
\item Temporal averaging at the scale of the whole musical note,
\item Median-based logarithmic compression,
\item Affine standardization so that each feature has zero mean and unit variance, and
\item Nearest-neighbor search according to previously learned non-Euclidean metric.
\end{enumerate}

Note that the non-Euclidean metric is learned via LMNN (see \lnameref{sec:methods} section) on the ``consensus'' cluster graph $\mathcal{G}_0$.
Therefore, the system $\mathbf{\Phi}$ performs timbre similarity retrieval in a user-agnostic way, and can serve as a convenient default algorithm for newcoming users.
That being said, it is conceivable to replicate the five-stage protocol above on the cluster graph $\mathcal{G}_k$ of a specific user $k$, instead of the cluster graph $\mathcal{G}_0$.
This operation would lead to a new configuration of the search engine $\mathbf{\Phi}$ that is better tailored to the perceptual idiosyncrasy of user $k$ in terms of timbre similarity.

Within the default setting ($\mathcal{G}=\mathcal{G}_0$), our system $\mathbf{\Phi}$ achieves an average precision at rank five (AP@5) of $\textbf{99.0\%}$, with a standard deviation across $K=31$ participants of the order of $1\%$.
This favorable result suggests that joint time--frequency scattering provides a useful feature map for learning similarities between instrumental playing techniques.
In doing so, it is in line with a recent publication \cite{wang2020icassp}, in which the authors successfully trained a supervised classifier on joint time--frequency scattering features in order to detect and classify playing techniques from the Chinese bamboo flute \emph{(dizi)}.
However, the originality of our work is that $\mathbf{\Phi}$ relies purely on auditory information (i.e., timbre similarity judgments), and does not require any supervision from the symbolic domain.
In particular, it does not assume the metadata (instrument, mute, technique, pitch, dynamics, and so forth) of any musical sample $\boldsymbol{x}_n$ to be observable, in part or in full, at training time.

\subsection*{Visualization of joint--time frequency scattering coefficients}
\label{sec:visualization}

In the second order, joint time--frequency scattering coefficients depend upon four variables: time $t$, log-frequency $\lambda$, temporal modulation rate $\alpha$ in Hertz, and frequential modulation rate $\beta$ in cycles per octave.
From a data visualization standpoint, rendering the four-dimensional tensor $\mathbf{S_2}\boldsymbol{x}(t, \lambda, \alpha, \beta)$ is impossible.
To address this limitation, a recent publication has projected this tensor into a two-dimensional ``slice'', thus yielding an image raster \cite{anden2019tsp}.
In accordance with their protocol, we compute the following matrix:

\begin{equation}
\mathbf{V}\boldsymbol{x}(\alpha, \beta) =
\iint_{\mathbb{R}^2}
\mathbf{U_2}\boldsymbol{x}(t - t^\prime, \log \lambda - u^\prime)
\;\mathrm{d}t^\prime \, \mathrm{d}u^\prime,
\end{equation}

Observe that the equation above is a limit case of $\mathbf{S_2}$ (Equation \ref{eq:S2}) in which the constants $T$ and $F$ tend towards infinity.
As a result, $\mathbf{V}\boldsymbol{x}$ depends solely upon scale $\alpha$ and rate $\beta$.
In the scientific literature on STRF, the matrix $\mathbf{V}\boldsymbol{x}$ is known as the ``cortical output collapsed on the rate-scale axes'' \cite{elhilali2003speechcomm}.

Previous publications on STRFs have demonstrated the interest of visualizing the slice $\mathbf{V}\boldsymbol{x}$ in the case of speech \cite{bellur2015ciss}, lung sounds \cite{emmanouilidou2012embs}, and music \cite{patil2012ploscompbiol}.
However, the visualization of musical sounds has been restricted to some of most common playing techniques, i.e., piano played staccato and violin played pizzicato.
Furthermore, prior publications on time--frequency scattering have displayed slices of $\mathbf{S_2}\boldsymbol{x}$ in the case of synthetic signals; but there is a gap in literature as regards the interpretability of the scale--rate domain in the case of real-world signals.

To remedy this gap, we select twelve isolated notes from the SOL dataset from two instruments: violin (Figure \ref{fig:violin-scattering}) and flute (Figure \ref{fig:flute-scattering}).
By and large, we find that joint time--frequency scattering produces comparable patterns in the scale--rate domain as the ``cortical output'' of the STRF.
For example, Figure \ref{fig:violin-scattering}(a) shows that a violin note played in the \emph{ordinario} technique has a local energy maximum at the rate $\alpha = \SI{6}{\hertz}$.
A visual inspection of $\mathbf{U_1}$ demonstrates that this rate coincides with the rate of vibrato of the left hand in the violin note.
As seen in Figure \ref{fig:violin-scattering}(b), this local energy maximum is absent when the playing technique is denoted as \emph{nonvibrato}.
Furthermore, Figure \ref{fig:violin-scattering}(c) shows that the local energy maximum is displaced to a higher rate ($\alpha = \SI{12}{\hertz}$) when the vibrato is replaced by a tremolo.

The visual interpretation of playing techniques in terms of their joint time--frequency scattering coefficients is not restricted to periodic modulation, such as vibrato or tremolo: rather, it also encompasses the analysis of attack transients.
Figures \ref{fig:violin-scattering}(d), (e), and (f) show the matrix $\mathbf{V}\boldsymbol{x}$ for three instances of impulsive violin sounds: sforzando, pizzicato, and staccato respectively.
These three techniques create ridges in the scale--rate domain $(\alpha, \beta)$, where the cutoff rate $\alpha$ is lowest with sforzando and highest with staccato.
These variations in cutoff rate coincide with perceptual variations in ``hardness'', i.e. impulsivity, of the violin sound.
Moreover, in the case of staccato, we observe a slight asymmetry in the frequential scale parameter $\beta$.
This asymmetry could be due to the fact that higher-order harmonics decay faster than the fundamental, thus yielding a triangular shape in the time--frequency domain.

Figure \ref{fig:flute-scattering} shows six playing techniques of the flute.
Similarly to the violin (Figure \ref{fig:violin-scattering}), we observe that periodic modulations, such as a trill (Figure \ref{fig:flute-scattering}(b)) or a beating tone (Figure \ref{fig:flute-scattering}(c)), cause local energy maxima whose rate $\alpha$ is physically interpretable.
Likewise, impulsive flute sounds such as sforzando (Figure \ref{fig:flute-scattering}(d)), key click (Figure \ref{fig:flute-scattering}(e)), and vibrato (Figure \ref{fig:flute-scattering}(f)) create ridges in the scale--rate domain of varying cutoff rates $\alpha$.
We distribute the implementation of these figures as part of the MATLAB library \emph{scattering.m}, which is released under the MIT license.\footnote{Link to download the \emph{scattering.m} library: \url{https://github.com/lostanlen/scattering.m}}

\subsection*{Ablation study}
We now turn to alter certain key choices in the design of the above-described computational blocks, and discuss their respective impacts on downstream performance.

Figure \ref{fig:ablation} summarizes our results.
Interestingly, the system $\boldsymbol{\Phi}$ is not only best on average, but also best for every subject in the cohort.
Specifically, replacing $\boldsymbol{\Phi}$ by a simpler model $\boldsymbol{\Phi}^{\prime}$ (see subsections below for examples of such models) results in $\mathrm{P}_{\mathbf{\Phi^\prime}}(k, 5) < \mathrm{P}_{\mathbf{\Phi}}(k, 5)$ for every $k$.
Borrowing from the terminology of welfare economics, $\boldsymbol{\Phi}$ can be said to be uniquely Pareto-efficient \cite{black2012dictionary}.
This observation suggests that the increase in performance afforded by the state-of-the-art model with respect to the baseline does not come at the detriment of user fairness.

\subsection*{Role of dataset size}
First of all, it is worth noting that the presented AP@5 figure of $99.0\% \pm 1$ does not abide to a conventional training set vs. test set paradigm, as is most often done in machine learning research.
Rather, the LMNN algorithm is trained on all available samples ($N^{\prime}=9346$ isolated notes) with the ``consensus'' cluster graph as training objective ($\mathcal{G}=\mathcal{G}_0$).
Then, it is evaluated on the same samples with individual cluster graphs as ground truth ($\mathcal{G}=\mathcal{G}_k$ for $k\geq1$).
The reason behind this choice is that, in the context of computer-assisted orchestration, the collection of audio samples has a fixed size, as it is shipped alongside the software itself.
This is the case, for example, of Orchidea,\footnote{Link to Orchidea software and OrchideaSOL dataset: \url{www.orch-idea.org}} which comes paired with its own version of SOL, named OrchideaSOL \cite{cella2020icmc}.
Henceforth, our main goal was to evaluate the generalization ability of our metric learning algorithm beyond the restricted set of samples for which human annotations are directly available (see \lnameref{sec:data-collection} section); that is, beyond one pitch class (middle C) and one intensity level (\emph{mf}).

Despite this caveat, we may adopt a ``query-by-example'' (QbE) paradigm by  partitioning the database and audio samples in half ($\frac{1}{2}N^{\prime}=4673$), training the LMNN algorithm on the first half, and querying it with samples from the other half.
In this evaluation framework, our system reaches an AP@5 of $\textbf{96.2\%} \pm 2$.
Interestingly, querying the system with samples from the training set leads to an AP@5 of $\textbf{96.5\%} \pm 2$, i.e., roughly on par with the test set.
Thus, it appears that the gap in performance between the evaluation presented in the \lnameref{sec:results} section ($99.0\%\pm 1$) and query-by-example evaluation ($96.2\% \pm 2$) is primarily attributable to a reduction of the size of training set by half, whereas statistical overfitting of the training set with respect to the test set likely plays a minor role.

These findings are in line with a previous publication \cite{patil2012ploscompbiol} which modeled perceived dissimilarity between musical sounds by means of STRF features.
Likewise, a recent publication has observed similar performance in instrument identification of a human and machine classifier using separable Gabor filterbank (GBFB) features \cite{siedenburg2019jasa}.
We note that both STRF and GBFB bear a strong computational resemblance with time--frequency scattering.


\subsection*{Role of metric learning}
Replacing the LMNN metric learning algorithm by linear discriminant analysis (LDA) leads to an AP@5 of $\textbf{76.6}\% \pm 11$.
Moreover, we evaluate the nearest neighbor algorithm $\mathbf{\Phi}$ in the absence of any metric learning at all. 
This corresponds to using a Euclidean distance to compare scattering coefficients, i.e., to set $\mathbf{L}$ to the identity matrix.
Note that the runtime complexity of Euclidean nearest-neighbor search is identical to LMNN search. 
We report an average precision at rank five (AP@5) of $\textbf{92.9\%} \pm 3$, which is noticeably worse than the best performing system.
This gap in performance demonstrates the importance of complementing unsupervised feature extraction by supervised metric learning in the design  of computational models for timbre similarity between instrumental playing techniques.

\subsection*{Role of temporal context $T$}

Our best performing system operates with joint time--frequency scattering coefficients as spectrotemporal modulation features.
These features are extracted within a temporal context of duration equal to $T=\SI{1000}{\milli\second}$.
This value is considerably larger than the frame size of purely spectral features, such as spectral centroid, spectral flux, or mel-frequency cepstral coefficients (MFCCs).
Indeed, the frame size of spectral features for machine listening is typically set to $T=\SI{23}{\milli\second}$, i.e., $2^{10}=1024$ samples at a sampling rate of $\SI{44,1}{\kilo\hertz}$ \cite{brown1999jasa,eronen2000icassp}.

As a point of comparison, we set the maximum time scale of joint time--frequency scattering coefficients to $T=\SI{25}{\milli\second}$, hence a $40$-fold reduction in context size.
Over our cohort of $K=31$ participants, we report an AP@5 of $\textbf{90.9\%} \pm 4$, which is noticeably worse than the best performing system.
This gap in performance extends the findings of a previous publication \cite{lostanlen2018extended}, which reported that metric learning with temporal scattering coefficients tends to improve with growing values of $T$, until reaching a plateau of performance around $T\sim\SI{500}{\milli\second}$.

\subsection*{Role of joint time--frequency scattering}

Let us recall the full definition of second-order joint time--frequency scattering (see \lnameref{sec:methods} section):
\begin{equation}
\mathbf{S_2}\boldsymbol{x}(t,\lambda,\alpha,\beta) =
\left(
\Big\vert
\big\vert
\boldsymbol{x} \ast \boldsymbol{\psi}_{\lambda}
\big\vert
\circledast
\boldsymbol{\Psi}_{\alpha,\beta}
\Big\vert
\circledast
\big(\boldsymbol{\phi}_{T} \otimes \boldsymbol{\phi}_{F})
\right)(t,\lambda),
\label{eq:joint-scattering}
\end{equation}
where the $\boldsymbol{\psi}_{\lambda}$ denotes a Morlet wavelet of center frequency $\lambda$ and resp. $\boldsymbol{\phi}_T$ denotes a Gaussian low-pass filter of width $T$.
Besides the \emph{joint} time--frequency scattering transform, the generative grammar of scattering transforms \cite{lostanlen2019chapter} also encompasses the \emph{separable} time--frequency scattering transform:
\begin{equation}
\mathbf{S_2^{\mathrm{sep}}}\boldsymbol{x}(t,\lambda,\alpha,\beta) =
\left(
\bigg\vert
\Big\vert
\big\vert
\boldsymbol{x} \ast \boldsymbol{\psi}_{\lambda}
\big\vert
\ast
\boldsymbol{\psi}_{\alpha}
\Big\vert
\circledast
\big(\boldsymbol{\phi}_{T} \otimes \boldsymbol{\psi}_{\beta}\big)
\bigg\vert
\circledast
\big(\boldsymbol{\phi}_{T} \otimes \boldsymbol{\phi}_F\big)
\right)(t, \lambda),
\label{eq:separable-scattering}
\end{equation}
where the wavelet $\boldsymbol{\psi}_{\lambda}$ has a quality factor of $Q=12$ whereas the wavelets $\boldsymbol{\psi}_{\alpha}$ and $\boldsymbol{\psi}_{\beta}$ have a quality factor of one.
Previous publications have successfully applied separable time--frequency scattering in order to classify environmental sounds \cite{bauge2013icassp} as well as playing techniques from the Chinese bamboo flute \cite{wang2019ismir}.

In comparison with its joint counterpart, separable time--frequency scattering contains about half as many coefficients.
This is because the temporal wavelet transform with $\boldsymbol{\psi}_\alpha$ and the frequential wavelet transform with $\boldsymbol{\psi}_\beta$ are separated by an operation of complex modulus.
Hence, $\boldsymbol{\psi}_\beta$ operates on a real-valued input.
Because separable time--frequency scattering cannot distinguish ascending chirps from descending chirps, flipping the sign of the scale variable $\beta$ is no longer necessary.
Moreover, separable time--frequency scattering has a lower algorithmic complexity than joint time--frequency scattering.
Indeed, in Equation \ref{eq:separable-scattering}, the frequential wavelet transform with $\boldsymbol{\psi}_{\beta}$ operates on a tensor whose time axis is subsampled at a fixed rate $T^{-1}$, thus allowing vectorized computations.
Conversely, in Equation \ref{eq:joint-scattering}, the frequential wavelet transform must operate on a multiresolution input, whose sample rate varies  depending on $\alpha$: it ranges between $T^{-1}$ and the sample rate of $\boldsymbol{x}$ itself.

Yet, despite its greater simplicity, separable time--frequency scattering suffers from known weaknesses in its ability to represent spectrotemporal modulations.
In particular, a previous publication has shown that frequency-dependent time shifts affect joint time--frequency scattering coefficients while leaving separable time--frequency scattering coefficients almost unchanged \cite{anden2019tsp}.
The same observation was made by \cite{schadler2015jasa} in the case of joint and separable Gabor filterbank features (GBFB), which bear some resemblance with joint and separable time--frequency scattering coefficients respectively.

Over our cohort of $K=31$ participants, separable time--frequency scattering achieves n AP@5 of $\textbf{91.9\%} \pm 4$. This figure is noticeably worse than joint time--frequency scattering ($\textbf{99.0}\% \pm 1$), all other things being equal.
This gap in performance, together with the theory of STRFs in auditory neuroscience (see \lnameref{sec:related-work} section), demonstrates the importance of joint spectrotemporal modulations in the modeling of timbre similarity across instrumental playing techniques.

\subsection*{Comparison with mel--frequency cepstral coefficient (MFCC) baseline}
Lastly, we train a baseline system in which joint time--frequency scattering coefficients are replaced by MFCCs.
Specifically, we extract a 40-band mel-frequency spectrogram by means of the RASTAMAT library; apply the pointwise logarithm; and compute a discrete cosine transform (DCT) over the mel-frequency axis.\footnote{Link to RASTAMAT library: \url{http://www.ee.columbia.edu/~dpwe/resources/matlab/rastamat}}
This operation results in a 40-dimensional feature vector over frames of duration $T=\SI{25}{\milli\second}$.
Over our cohort of $K=31$ participants, we report an AP@5 of $\textbf{81.8\%} \pm 7$.

Arguably, this figure is not directly comparable with our best performing system (AP@5 of $99.0\%\pm 1$), due to the mismatch in dimensionality between MFCC and joint time--frequency scattering coefficients.
In order to clarify the role of feature dimensionality in our computational pipeline, we apply a feature engineering technique involving multiplicative combinations of MFCC.
We construct the following Gram matrix:
\begin{equation}
\mathbf{G} \boldsymbol{x}(\alpha, \beta) =
\int_0^{+\infty}
\mathrm{MFCC}(\boldsymbol{x})(t, \alpha)
\mathrm{MFCC}(\boldsymbol{x})(t, \beta)
\;\mathrm{d}t,
\end{equation}
where $\alpha$ and $\beta$ represent different dimensions (``quefrencies'') of the MFCC feature vector.
The symmetric matrix $\mathbf{G}\boldsymbol{x}$ contains $40$ rows and $40$ columns, hence $800$ unique coefficients.
Concatenating these coefficients to the $40$ averaged MFCC features results in a feature vector of $840$ coefficients.
This dimension is of the same order of magnitude as the dimension of our joint time--frequency scattering representation ($d=1180$, see \lnameref{sec:methods} section).

Training the LMNN algorithm on this 840-dimensional representation is analogous to a ``kernel trick'' in support vector machines \cite{chang2010jmlr}.
In our case, the implicit similarity kernel is a homogeneous quadratic kernel.
Despite this increase in representational power, we obtain an AP@5 of $81.5\% \pm 7$, i.e. essentially the same as MFCC under a linear kernel.
Therefore, it appears that the gap in performance between MFCC ($81.8\%\pm 7$) and joint time--frequency scattering ($99.0\% \pm 1$) is primarily attributable to the multiscale extraction of joint spectrotemporal modulations, whereas high-dimensional embedding likely plays a minor role.


\section*{\nmu Conclusion}
\label{sec:conclusion}

We see from the ablation study conducted above that each step of the proposed model is necessary for its high performance.
Among other things, it indicates the joint time--frequency scattering transform itself should not be used directly for similarity comparisons, but is best augmented by a learned stage.
This explains the relative success of such models \cite{lostanlen2018jasmp,lostanlen2018extended} over others where distances on raw scattering coefficients were used for judging audio similarity.

On the other hand, we note that the complexity of the learned model does not need to be high.
Indeed, for this task, a linear model on the scattering coefficients is sufficient.
There is no need for deep networks with large numbers of parameters to accurately represent the similarity information.
In other words, the scattering transform parametrizes the signal structure in a way that many relevant quantities (such as timbre similarity) can be extracted through a simple linear mapping.
This is in line with other classification results, where linear support vector machines applied to scattering coefficients have achieved significant success \cite{anden2015mlsp,anden2019tsp}.

We also see the necessity of a fully \emph{joint} time--frequency model for accessing timbre, as opposed to a purely spectral model or one that treats the two axes in a separable manner.
This fact has also been observed in other contexts, such as the work of Patil et al. \cite{patil2012ploscompbiol}.
A related observation is the need for capturing large-scale structure.
Indeed, reducing the window size to $25~\mathrm{ms}$ means that we lose a great deal of time--frequency structure, bringing results closer to that of the separable model.

The success of the above system in identifying timbral similarities has immediate applications in browsing music databases.
These are typically organized based on instrumental and playing techniques taxonomies, with additional keywords offering a more flexible organization.
Accessing the sounds in these databases therefore requires some knowledge of the taxonomy and keywords used.
Furthermore, the user needs to have some idea of the particular playing technique they are searching for.

As an alternative, content-based searches allow the user to identify sounds based on similarity to some given query sound.
This query-by-example approach provides an opportunity to search for sounds without having a specific instrument of playing technique in mind, yielding a wider range of available sounds.
A composer with access to such a system would therefore be able to draw on a more diverse palette of musical timbre.

The computational model proposed in this work is well suited to such a query-by-example task.
We have shown that it is able to adequately approximate the timbral judgments of a wide range of participants included in our study.
Not only that, but the system can be easily retrained to approximate an individual user's timbre perception by having that user perform the clustering task on the reduced set of $78$ IPTs and running the LMNN training step on those clustering assignments.
This model can then be applied to new data, or, alternatively, be retrained with these new examples if the existing model proves unsatisfactory.

We shall note, however, that the current model has several drawbacks.
First, it is only applied to instrumental sounds.
While this has the advantage of simplifying the interpretation of the results, the range of timbre under consideration is necessarily limited (although less restricted than only considering \emph{ordinario} PTs).
This also makes applications such as \emph{query-by-humming} difficult, since we cannot guarantee that the timbral similarity measure is accurate for such sounds.

That being said, the above model is general enough to encompass a wide variety of recordings, not just instrumental sounds.
Indeed, we have strong assumptions that the tools used (scattering transforms and LMNN weighting matrices) do not depend strongly on the type of sound being processed. Future work will investigate whether more general classes of sounds are also well modeled.
To extend the model, it is only necessary to retrain the LMNN weighting matrix by supplying it with new cluster assignments.
These can again be obtained by performing a new clustering experiment with one or more human subjects.

Another aspect is the granularity of the similarity judgments.
In the above method, we have used \emph{hard} clustering assignments to build our model.
A more nuanced similarity judgment would ask users to rate the similarity of a pair of IPTs on a more graduated scale, which would yield a finer, or \emph{soft}, assignment.
This however, comes with additional difficulties in providing a consistent scale across participants, but could be feasible if the goal is to only adapt the timbral space to a single individual.
An approach not based on clustering would also have to replace the LMNN algorithm with one that accounts for such soft assignments.

\begin{backmatter}

\section*{Abbreviations}

\begin{itemize}
  \item AP@5: average precision at rank 5
  \item CNSMDP: conservatoire national sup\'erieur de musique et de danse de Paris
  \item COSMIR: community for open and sustainable music information research
  \item DCT: discrete cosine transform
  \item FMA: free music archive
  \item GBFB: Gabor filterbank features
  \item IMT: instrument--mute--technique
  \item LMNN: large-margin nearest neighbor
  \item MFCC: mel-frequency cepstral coefficients
  \item MIR: music information retrieval
  \item P@5: precision at rank five
  \item ReLU: rectified linear unit
  \item STRF: spectrotemporal receptive fields
  \item SOL: Studio On Line dataset
\end{itemize}

\subsection*{Musical instruments}

\begin{itemize}
  \item Vn: violin,
  \item Va: viola,
  \item Vc: cello,
  \item Cb: contrabass,
  \item Hp: concert harp,
  \item Gtr: Spanish guitar,
  \item Acc: accordion,
  \item Fl: flute,
  \item BbCl: soprano clarinet,
  \item ASax: alto saxophone,
  \item Ob: oboe,
  \item Bn: bassoon,
  \item TpC: trumpet in C,
  \item Hn: French horn,
  \item TTbn: tenor trombone,
  \item BBTb: bass tuba.
\end{itemize}

\subsection*{Musical nuances}

\begin{itemize}
  \item \emph{pp}: pianissimo,
  \item \emph{p}: piano,
  \item \emph{mf}: mezzo forte,
  \item \emph{f}: forte,
  \item \emph{ff}: fortissimo.
\end{itemize}

\section*{Declarations}

\subsection*{\nmu Availability of data and materials}

The processing code, the Cyberlioz interface and an anonymized version of the  perceptual judgments gathered using this interface are available at the companion website: \url{https://github.com/mathieulagrange/lostanlen2020jasmp}.

The SOL dataset is available online: \url{https://forum.ircam.fr/projects/detail/fullsol}

\subsection*{\nmu Competing interests}
  The authors declare that they have no competing interests.


\subsection*{\nmu Funding}
This work is partially supported by the Paris sciences et lettres (PSL) TICEL project.
This work is partially supported by the European Research Council (ERC) award 320959 (InvariantClass).
This work is partially supported by the National Science Foundation (NSF) award 1633259 (BIRDVOX).
This work is partially supported by the Flatiron Institute, a division of the Simons Foundation.

\subsection*{\nmu Acknowledgments}

We wish to thank Philippe Brandeis, \'{E}tienne Graindorge, St\'{e}phane Mallat, Adrien Mamou-Mani, and Yan Maresz for contributing to the TICEL research project.
We also wish to thank the students of the Paris Conservatory and all anonymous participants to our study.

\subsection*{Authors' contributions}

VL provided guidance in the design of the computational experiments and was a major contributor in writing the manuscript. CEH conducted part of the computational experiments. MR designed and implemented the listening tests. GL designed the listening tests and provided guidance in the design of the computational experiments. JA provided guidance in the design of the computational experiments, participated to the data analysis and  to the writing of the manuscript. ML designed and conducted the computational experiments, the data analysis and participated to the writing of the paper.



\newcommand{\BMCxmlcomment}[1]{}

\BMCxmlcomment{

<refgrp>

<bibl id="B1">
  <title><p>Music information retrieval</p></title>
  <aug>
    <au><snm>Downie</snm><fnm>JS</fnm></au>
  </aug>
  <source>Annual review of information science and technology</source>
  <publisher>Wiley Online Library</publisher>
  <pubdate>2003</pubdate>
  <volume>37</volume>
  <issue>1</issue>
  <fpage>295</fpage>
  <lpage>-340</lpage>
</bibl>

<bibl id="B2">
  <title><p>The Present, Past, and Future of Timbre Research</p></title>
  <aug>
    <au><snm>Siedenburg</snm><fnm>K</fnm></au>
    <au><snm>Saitis</snm><fnm>C</fnm></au>
    <au><snm>McAdams</snm><fnm>S</fnm></au>
  </aug>
  <source>Timbre: Acoustics, Perception, and Cognition</source>
  <publisher>Cham: Springer International Publishing</publisher>
  <editor>Siedenburg, Kai and Saitis, Charalampos and McAdams, Stephen and
  Popper, Arthur N. and Fay, Richard R.</editor>
  <pubdate>2019</pubdate>
  <fpage>1</fpage>
  <lpage>-19</lpage>
</bibl>

<bibl id="B3">
  <title><p>Verbal correlates of perceptual dimensions of timbre</p></title>
  <aug>
    <au><snm>Faure</snm><fnm>A</fnm></au>
    <au><snm>McAdams</snm><fnm>S</fnm></au>
    <au><snm>Nosulenko</snm><fnm>V</fnm></au>
  </aug>
  <source>Proceedings of the International Conference on Music Perception and
  Cognition (ICMPC)</source>
  <pubdate>1996</pubdate>
</bibl>

<bibl id="B4">
  <title><p>Extended playing techniques: the next milestone in musical
  instrument recognition</p></title>
  <aug>
    <au><snm>Lostanlen</snm><fnm>V</fnm></au>
    <au><snm>And{\'e}n</snm><fnm>J</fnm></au>
    <au><snm>Lagrange</snm><fnm>M</fnm></au>
  </aug>
  <source>Proceedings of the International Conference on Digital Libraries for
  Musicology (DLfM)</source>
  <pubdate>2018</pubdate>
  <fpage>1</fpage>
  <lpage>-10</lpage>
</bibl>

<bibl id="B5">
  <title><p>Musical Acoustics, Timbre, and Computer-Aided Orchestration
  Challenges</p></title>
  <aug>
    <au><snm>Antoine</snm><fnm>A</fnm></au>
    <au><snm>Miranda</snm><fnm>ER</fnm></au>
  </aug>
  <source>Proc. ISMA</source>
  <pubdate>2018</pubdate>
</bibl>

<bibl id="B6">
  <title><p>Knowledge Representation Issues in Musical Instrument Ontology
  Design.</p></title>
  <aug>
    <au><snm>Kolozali</snm><fnm>S</fnm></au>
    <au><snm>Barthet</snm><fnm>M</fnm></au>
    <au><snm>Fazekas</snm><fnm>G</fnm></au>
    <au><snm>Sandler</snm><fnm>MB</fnm></au>
  </aug>
  <source>Proceedings of the International Society on Music Information
  Retrieval (ISMIR) Conference</source>
  <pubdate>2011</pubdate>
</bibl>

<bibl id="B7">
  <title><p>Understanding optical music recognition</p></title>
  <aug>
    <au><snm>Calvo Zaragoza</snm><fnm>J</fnm></au>
    <au><snm>Haji\v{c} Jr.</snm><fnm>J</fnm></au>
    <au><snm>Pacha</snm><fnm>A</fnm></au>
  </aug>
  <source>ACM Computing Surveys</source>
  <pubdate>2020</pubdate>
</bibl>

<bibl id="B8">
  <title><p>Sound structure in music</p></title>
  <aug>
    <au><snm>Erickson</snm><fnm>R</fnm></au>
  </aug>
  <publisher>University of California Press</publisher>
  <pubdate>1975</pubdate>
</bibl>

<bibl id="B9">
  <title><p>Human dissimilarity ratings of musical instrument timbre: A
  computational meta-analysis</p></title>
  <aug>
    <au><snm>Thoret</snm><fnm>E</fnm></au>
    <au><snm>Caramiaux</snm><fnm>B</fnm></au>
    <au><snm>Depalle</snm><fnm>P</fnm></au>
    <au><snm>McAdams</snm><fnm>S</fnm></au>
  </aug>
  <source>Journal of the Acoustical Society of America</source>
  <publisher>ASA</publisher>
  <pubdate>2018</pubdate>
  <volume>143</volume>
  <issue>3</issue>
  <fpage>1745</fpage>
  <lpage>-1746</lpage>
</bibl>

<bibl id="B10">
  <title><p>On computer-assisted orchestration</p></title>
  <aug>
    <au><snm>Maresz</snm><fnm>Y</fnm></au>
  </aug>
  <source>Contemporary Music Review</source>
  <pubdate>2013</pubdate>
  <volume>32</volume>
  <issue>1</issue>
  <fpage>99</fpage>
  <lpage>-109</lpage>
</bibl>

<bibl id="B11">
  <title><p>Leveraging diversity in computer-aided musical orchestration with
  an artificial immune system for multi-modal optimization</p></title>
  <aug>
    <au><snm>Caetano</snm><fnm>M</fnm></au>
    <au><snm>Zacharakis</snm><fnm>A</fnm></au>
    <au><snm>Barbancho</snm><fnm>I</fnm></au>
    <au><snm>Tard{\'o}n</snm><fnm>LJ</fnm></au>
  </aug>
  <source>Swarm and Evolutionary Computation</source>
  <publisher>Elsevier</publisher>
  <pubdate>2019</pubdate>
  <volume>50</volume>
  <fpage>100484</fpage>
</bibl>

<bibl id="B12">
  <title><p>Joint time-frequency scattering for audio
  classification</p></title>
  <aug>
    <au><snm>And{\'e}n</snm><fnm>J</fnm></au>
    <au><snm>Lostanlen</snm><fnm>V</fnm></au>
    <au><snm>Mallat</snm><fnm>S</fnm></au>
  </aug>
  <source>Proceedings of the IEEE International Workshop on Machine Learning
  for Signal Processing (MLSP)</source>
  <pubdate>2015</pubdate>
  <fpage>1</fpage>
  <lpage>-6</lpage>
</bibl>

<bibl id="B13">
  <title><p>Joint Time--Frequency Scattering</p></title>
  <aug>
    <au><snm>And\'{e}n</snm><fnm>J</fnm></au>
    <au><snm>Lostanlen</snm><fnm>V</fnm></au>
    <au><snm>Mallat</snm><fnm>S</fnm></au>
  </aug>
  <source>IEEE Transactions on Signal Processing</source>
  <pubdate>2019</pubdate>
  <volume>67</volume>
  <issue>14</issue>
  <fpage>3704</fpage>
  <lpage>-3718</lpage>
</bibl>

<bibl id="B14">
  <title><p>Relevance-based quantization of scattering features for
  unsupervised mining of environmental audio</p></title>
  <aug>
    <au><snm>Lostanlen</snm><fnm>V</fnm></au>
    <au><snm>Lafay</snm><fnm>G</fnm></au>
    <au><snm>And{\'e}n</snm><fnm>J</fnm></au>
    <au><snm>Lagrange</snm><fnm>M</fnm></au>
  </aug>
  <source>EURASIP Journal on Audio, Speech, and Music Processing</source>
  <pubdate>2018</pubdate>
  <volume>2018</volume>
  <issue>1</issue>
  <fpage>15</fpage>
</bibl>

<bibl id="B15">
  <title><p>Scattering Representation of Modulated Sounds</p></title>
  <aug>
    <au><snm>And\'{e}n</snm><fnm>J</fnm></au>
    <au><snm>Mallat</snm><fnm>S</fnm></au>
  </aug>
  <source>Proceedings of the International Conference on Digital Audio Effects
  (DAFx)</source>
  <pubdate>2012</pubdate>
</bibl>

<bibl id="B16">
  <title><p>Distance metric learning for large margin nearest neighbor
  classification</p></title>
  <aug>
    <au><snm>Weinberger</snm><fnm>KQ</fnm></au>
    <au><snm>Saul</snm><fnm>LK</fnm></au>
  </aug>
  <source>Journal of Machine Learning Research</source>
  <pubdate>2009</pubdate>
  <volume>10</volume>
  <issue>Feb</issue>
  <fpage>207</fpage>
  <lpage>-244</lpage>
</bibl>

<bibl id="B17">
  <title><p>Perceptual scaling of synthesized musical timbres: common
  dimensions, specificities, and latent subject classes</p></title>
  <aug>
    <au><snm>McAdams</snm><fnm>S</fnm></au>
    <au><snm>Winsberg</snm><fnm>S</fnm></au>
    <au><snm>Donnadieu</snm><fnm>S</fnm></au>
    <au><snm>De Soete</snm><fnm>G</fnm></au>
    <au><snm>Krimphoff</snm><fnm>J</fnm></au>
  </aug>
  <source>Psychological research</source>
  <publisher>Springer</publisher>
  <pubdate>1995</pubdate>
  <volume>58</volume>
  <issue>3</issue>
  <fpage>177</fpage>
  <lpage>-192</lpage>
</bibl>

<bibl id="B18">
  <title><p>Music in our ears: {The} biological bases of musical timbre
  perception</p></title>
  <aug>
    <au><snm>Patil</snm><fnm>K</fnm></au>
    <au><snm>Pressnitzer</snm><fnm>D</fnm></au>
    <au><snm>Shamma</snm><fnm>S</fnm></au>
    <au><snm>Elhilali</snm><fnm>M</fnm></au>
  </aug>
  <source>PLoS Computational Biology</source>
  <publisher>Public Library of Science</publisher>
  <pubdate>2012</pubdate>
  <volume>8</volume>
  <issue>11</issue>
  <fpage>e1002759</fpage>
</bibl>

<bibl id="B19">
  <title><p>Temporal integration for audio classification with application to
  musical instrument classification</p></title>
  <aug>
    <au><snm>Joder</snm><fnm>C</fnm></au>
    <au><snm>Essid</snm><fnm>S</fnm></au>
    <au><snm>Richard</snm><fnm>G</fnm></au>
  </aug>
  <source>IEEE Transactions on Audio, Speech, and Language Processing</source>
  <publisher>IEEE</publisher>
  <pubdate>2009</pubdate>
  <volume>17</volume>
  <issue>1</issue>
  <fpage>174</fpage>
  <lpage>-186</lpage>
</bibl>

<bibl id="B20">
  <title><p>A comparison of approaches to timbre descriptors in music
  information retrieval and music psychology</p></title>
  <aug>
    <au><snm>Siedenburg</snm><fnm>K</fnm></au>
    <au><snm>Fujinaga</snm><fnm>I</fnm></au>
    <au><snm>McAdams</snm><fnm>S</fnm></au>
  </aug>
  <source>Journal of New Music Research</source>
  <publisher>Taylor \& Francis</publisher>
  <pubdate>2016</pubdate>
  <volume>45</volume>
  <issue>1</issue>
  <fpage>27</fpage>
  <lpage>-41</lpage>
</bibl>

<bibl id="B21">
  <title><p>Musical instrument identification: A pattern recognition
  approach</p></title>
  <aug>
    <au><snm>Martin</snm><fnm>KD</fnm></au>
    <au><snm>Kim</snm><fnm>YE</fnm></au>
  </aug>
  <source>Proceedings of the Acoustical Society of America</source>
  <pubdate>1998</pubdate>
</bibl>

<bibl id="B22">
  <title><p>Computer identification of musical instruments using pattern
  recognition with cepstral coefficients as features</p></title>
  <aug>
    <au><snm>Brown</snm><fnm>JC</fnm></au>
  </aug>
  <source>The Journal of the Acoustical Society of America</source>
  <publisher>ASA</publisher>
  <pubdate>1999</pubdate>
  <volume>105</volume>
  <issue>3</issue>
  <fpage>1933</fpage>
  <lpage>-1941</lpage>
</bibl>

<bibl id="B23">
  <title><p>Musical instrument recognition using cepstral coefficients and
  temporal features</p></title>
  <aug>
    <au><snm>Eronen</snm><fnm>A</fnm></au>
    <au><snm>Klapuri</snm><fnm>A</fnm></au>
  </aug>
  <source>Proceedings of the IEEE International Conference on Acoustics,
  Speech, and Signal Processing (ICASSP)</source>
  <pubdate>2000</pubdate>
</bibl>

<bibl id="B24">
  <title><p>Automatic classification of musical instrument sounds</p></title>
  <aug>
    <au><snm>Herrera Boyer</snm><fnm>P</fnm></au>
    <au><snm>Peeters</snm><fnm>G</fnm></au>
    <au><snm>Dubnov</snm><fnm>S</fnm></au>
  </aug>
  <source>Journal of New Music Research</source>
  <publisher>Taylor \& Francis</publisher>
  <pubdate>2003</pubdate>
  <volume>32</volume>
  <issue>1</issue>
  <fpage>3</fpage>
  <lpage>-21</lpage>
</bibl>

<bibl id="B25">
  <title><p>Analysis of feature dependencies in sound description</p></title>
  <aug>
    <au><snm>Wieczorkowska</snm><fnm>AA</fnm></au>
    <au><snm>{\.Z}ytkow</snm><fnm>JM</fnm></au>
  </aug>
  <source>Journal of Intelligent Information Systems</source>
  <publisher>Springer</publisher>
  <pubdate>2003</pubdate>
  <volume>20</volume>
  <issue>3</issue>
  <fpage>285</fpage>
  <lpage>-302</lpage>
</bibl>

<bibl id="B26">
  <title><p>Musical instrument identification in continuous
  recordings</p></title>
  <aug>
    <au><snm>Livshin</snm><fnm>A</fnm></au>
    <au><snm>Rodet</snm><fnm>X</fnm></au>
  </aug>
  <source>Proceedings of the International Conference on Digital Audio Effects
  (DAFx)</source>
  <pubdate>2004</pubdate>
</bibl>

<bibl id="B27">
  <title><p>Music instrument recognition: from isolated notes to solo
  phrases</p></title>
  <aug>
    <au><snm>Krishna</snm><fnm>A.G.</fnm></au>
    <au><snm>Sreenivas</snm><fnm>TV</fnm></au>
  </aug>
  <source>Proceedings of the IEEE International Conference on Acoustics,
  Speech, and Signal Processing (ICASSP)</source>
  <pubdate>2004</pubdate>
</bibl>

<bibl id="B28">
  <title><p>{Automatic recognition of isolated monophonic musical instrument
  sounds using kNNC}</p></title>
  <aug>
    <au><snm>Kaminskyj</snm><fnm>I</fnm></au>
    <au><snm>Czaszejko</snm><fnm>T</fnm></au>
  </aug>
  <source>Journal of Intelligent Information Systems</source>
  <publisher>Springer</publisher>
  <pubdate>2005</pubdate>
  <volume>24</volume>
  <issue>2-3</issue>
  <fpage>199</fpage>
  <lpage>-221</lpage>
</bibl>

<bibl id="B29">
  <title><p>Musical instrument classification using non-negative matrix
  factorization algorithms and subset feature selection</p></title>
  <aug>
    <au><snm>Benetos</snm><fnm>E</fnm></au>
    <au><snm>Kotti</snm><fnm>M</fnm></au>
    <au><snm>Kotropoulos</snm><fnm>C</fnm></au>
  </aug>
  <source>Proceedings of the IEEE International Conference on Acoustics,
  Speech, and Signal Processing (ICASSP)</source>
  <pubdate>2006</pubdate>
</bibl>

<bibl id="B30">
  <title><p>{Automatic musical instrument classification using fractional
  Fourier transform based-MFCC features and counter propagation neural
  network}</p></title>
  <aug>
    <au><snm>Bhalke</snm><fnm>D.G.</fnm></au>
    <au><snm>Rao</snm><fnm>CR</fnm></au>
    <au><snm>Bormane</snm><fnm>DS</fnm></au>
  </aug>
  <source>Journal of Intelligent Information Systems</source>
  <publisher>Springer</publisher>
  <pubdate>2016</pubdate>
  <volume>46</volume>
  <issue>3</issue>
  <fpage>425</fpage>
  <lpage>-446</lpage>
</bibl>

<bibl id="B31">
  <title><p>OpenMIC-2018: an open dataset for multiple instrument
  recognition</p></title>
  <aug>
    <au><snm>Humphrey</snm><fnm>E</fnm></au>
    <au><snm>Durand</snm><fnm>S</fnm></au>
    <au><snm>McFee</snm><fnm>B</fnm></au>
  </aug>
  <source>Proceedings of the International Society on Music Information
  Retrieval (ISMIR) Conference</source>
  <pubdate>2018</pubdate>
</bibl>

<bibl id="B32">
  <title><p>{A plan for sustainable {MIR} evaluation}</p></title>
  <aug>
    <au><snm>McFee</snm><fnm>B</fnm></au>
    <au><snm>Humphrey</snm><fnm>EJ</fnm></au>
    <au><snm>Urbano</snm><fnm>J</fnm></au>
  </aug>
  <source>Proceedings of the International Society on Music Information
  Retrieval (ISMIR) Conference</source>
  <pubdate>2016</pubdate>
</bibl>

<bibl id="B33">
  <title><p>{FMA: A dataset for music analysis}</p></title>
  <aug>
    <au><snm>Defferrard</snm><fnm>M</fnm></au>
    <au><snm>Benzi</snm><fnm>K</fnm></au>
    <au><snm>Vandergheynst</snm><fnm>P</fnm></au>
    <au><snm>Bresson</snm><fnm>X</fnm></au>
  </aug>
  <source>Proceedings of the International Society on Music Information
  Retrieval (ISMIR) Conference</source>
  <pubdate>2017</pubdate>
</bibl>

<bibl id="B34">
  <title><p>Deep convolutional networks on the pitch spiral for musical
  instrument recognition</p></title>
  <aug>
    <au><snm>Lostanlen</snm><fnm>V</fnm></au>
    <au><snm>Cella</snm><fnm>CE</fnm></au>
  </aug>
  <source>Proceedings of the International Society on Music Information
  Retrieval (ISMIR) Conference</source>
  <pubdate>2016</pubdate>
</bibl>

<bibl id="B35">
  <title><p>{MedleyDB: A multitrack dataset for annotation-intensive MIR
  research}</p></title>
  <aug>
    <au><snm>Bittner</snm><fnm>RM</fnm></au>
    <au><snm>Salamon</snm><fnm>J</fnm></au>
    <au><snm>Tierney</snm><fnm>M</fnm></au>
    <au><snm>Mauch</snm><fnm>M</fnm></au>
    <au><snm>Cannam</snm><fnm>C</fnm></au>
    <au><snm>Bello</snm><fnm>JP</fnm></au>
  </aug>
  <source>Proceedings of the International Society on Music Information
  Retrieval (ISMIR) Conference</source>
  <pubdate>2014</pubdate>
</bibl>

<bibl id="B36">
  <title><p>A software framework for musical data augmentation</p></title>
  <aug>
    <au><snm>McFee</snm><fnm>B</fnm></au>
    <au><snm>Humphrey</snm><fnm>EJ</fnm></au>
    <au><snm>Bello</snm><fnm>JP</fnm></au>
  </aug>
  <source>Proceedings of the International Society on Music Information
  Retrieval (ISMIR)</source>
  <pubdate>2015</pubdate>
</bibl>

<bibl id="B37">
  <title><p>Timbre analysis of music audio signals with convolutional neural
  networks</p></title>
  <aug>
    <au><snm>{Pons}</snm><fnm>J.</fnm></au>
    <au><snm>{Slizovskaia}</snm><fnm>O.</fnm></au>
    <au><snm>{Gong}</snm><fnm>R.</fnm></au>
    <au><snm>{G\'{o}mez}</snm><fnm>E.</fnm></au>
    <au><snm>{Serra}</snm><fnm>X.</fnm></au>
  </aug>
  <source>25th European Signal Processing Conference (EUSIPCO)</source>
  <pubdate>2017</pubdate>
  <fpage>2744</fpage>
  <lpage>-2748</lpage>
</bibl>

<bibl id="B38">
  <title><p>Instrument Activity Detection in Polyphonic Music using Deep Neural
  Networks.</p></title>
  <aug>
    <au><snm>Gururani</snm><fnm>S</fnm></au>
    <au><snm>Summers</snm><fnm>C</fnm></au>
    <au><snm>Lerch</snm><fnm>A</fnm></au>
  </aug>
  <source>Proceedings of the International Society on Music Information
  Retrieval (ISMIR) Conference</source>
  <pubdate>2018</pubdate>
</bibl>

<bibl id="B39">
  <title><p>Timbre Classification Of A Single Musical Instrument</p></title>
  <aug>
    <au><snm>Loureiro</snm><fnm>MA</fnm></au>
    <au><snm>Paula</snm><fnm>HB</fnm></au>
    <au><snm>Yehia</snm><fnm>HC</fnm></au>
  </aug>
  <source>Proceedings of the International Society on Music Information
  Retrieval (ISMIR) Conference</source>
  <pubdate>2004</pubdate>
</bibl>

<bibl id="B40">
  <title><p>Deep convolutional neural networks for predominant instrument
  recognition in polyphonic music</p></title>
  <aug>
    <au><snm>Han</snm><fnm>Y</fnm></au>
    <au><snm>Kim</snm><fnm>J</fnm></au>
    <au><snm>Lee</snm><fnm>K</fnm></au>
  </aug>
  <source>IEEE Transactions on Audio, Speech, and Language Processing</source>
  <publisher>IEEE Press</publisher>
  <pubdate>2017</pubdate>
  <volume>25</volume>
  <issue>1</issue>
  <fpage>208</fpage>
  <lpage>-221</lpage>
</bibl>

<bibl id="B41">
  <title><p>The perception of musical timbre</p></title>
  <aug>
    <au><snm>McAdams</snm><fnm>S</fnm></au>
    <au><snm>Giordano</snm><fnm>BL</fnm></au>
  </aug>
  <source>The {O}xford handbook of music psychology</source>
  <pubdate>2009</pubdate>
  <fpage>72</fpage>
  <lpage>-80</lpage>
</bibl>

<bibl id="B42">
  <title><p>Acoustic and Categorical Dissimilarity of Musical Timbre: Evidence
  from Asymmetries Between Acoustic and Chimeric Sounds</p></title>
  <aug>
    <au><snm>Siedenburg</snm><fnm>K</fnm></au>
    <au><snm>Jones Mollerup</snm><fnm>K</fnm></au>
    <au><snm>McAdams</snm><fnm>S</fnm></au>
  </aug>
  <source>Frontiers in Psychology</source>
  <pubdate>2016</pubdate>
  <volume>6</volume>
  <fpage>1977</fpage>
</bibl>

<bibl id="B43">
  <title><p>Spectro-temporal response field characterization with dynamic
  ripples in ferret primary auditory cortex</p></title>
  <aug>
    <au><snm>Depireux</snm><fnm>DA</fnm></au>
    <au><snm>Simon</snm><fnm>JZ</fnm></au>
    <au><snm>Klein</snm><fnm>DJ</fnm></au>
    <au><snm>Shamma</snm><fnm>SA</fnm></au>
  </aug>
  <source>Journal of Neurophysiology</source>
  <publisher>Am Physiological Soc</publisher>
  <pubdate>2001</pubdate>
  <volume>85</volume>
  <issue>3</issue>
  <fpage>1220</fpage>
  <lpage>-1234</lpage>
</bibl>

<bibl id="B44">
  <title><p>The spectro-temporal receptive field</p></title>
  <aug>
    <au><snm>Aertsen</snm><fnm>AMHJ</fnm></au>
    <au><snm>Johannesma</snm><fnm>PIM</fnm></au>
  </aug>
  <source>Biological Cybernetics</source>
  <publisher>Springer</publisher>
  <pubdate>1981</pubdate>
  <volume>42</volume>
  <issue>2</issue>
  <fpage>133</fpage>
  <lpage>-143</lpage>
</bibl>

<bibl id="B45">
  <title><p>Triggered correlation</p></title>
  <aug>
    <au><snm>De Boer</snm><fnm>E</fnm></au>
    <au><snm>Kuyper</snm><fnm>P</fnm></au>
  </aug>
  <source>IEEE Transactions on Biomedical Engineering</source>
  <publisher>IEEE</publisher>
  <pubdate>1968</pubdate>
  <issue>3</issue>
  <fpage>169</fpage>
  <lpage>-179</lpage>
</bibl>

<bibl id="B46">
  <title><p>Time-frequency/time-scale analysis</p></title>
  <aug>
    <au><snm>Flandrin</snm><fnm>P</fnm></au>
  </aug>
  <publisher>Academic press</publisher>
  <pubdate>1998</pubdate>
</bibl>

<bibl id="B47">
  <title><p>Wiener and Volterra analyses applied to the auditory
  system</p></title>
  <aug>
    <au><snm>Eggermont</snm><fnm>JJ</fnm></au>
  </aug>
  <source>Hearing research</source>
  <publisher>Elsevier</publisher>
  <pubdate>1993</pubdate>
  <volume>66</volume>
  <issue>2</issue>
  <fpage>177</fpage>
  <lpage>-201</lpage>
</bibl>

<bibl id="B48">
  <title><p>Robust spectrotemporal reverse correlation for the auditory system:
  optimizing stimulus design</p></title>
  <aug>
    <au><snm>Klein</snm><fnm>DJ</fnm></au>
    <au><snm>Depireux</snm><fnm>DA</fnm></au>
    <au><snm>Simon</snm><fnm>JZ</fnm></au>
    <au><snm>Shamma</snm><fnm>SA</fnm></au>
  </aug>
  <source>Journal of Computational Neuroscience</source>
  <pubdate>2000</pubdate>
  <volume>9</volume>
  <issue>1</issue>
  <fpage>85</fpage>
  <lpage>-111</lpage>
</bibl>

<bibl id="B49">
  <title><p>Spectral-temporal receptive fields of nonlinear auditory neurons
  obtained using natural sounds</p></title>
  <aug>
    <au><snm>Theunissen</snm><fnm>FE</fnm></au>
    <au><snm>Sen</snm><fnm>K</fnm></au>
    <au><snm>Doupe</snm><fnm>AJ</fnm></au>
  </aug>
  <source>Journal of Neuroscience</source>
  <publisher>Soc. Neuroscience</publisher>
  <pubdate>2000</pubdate>
  <volume>20</volume>
  <issue>6</issue>
  <fpage>2315</fpage>
  <lpage>-2331</lpage>
</bibl>

<bibl id="B50">
  <title><p>Multiresolution spectrotemporal analysis of complex
  sounds</p></title>
  <aug>
    <au><snm>Chi</snm><fnm>T</fnm></au>
    <au><snm>Ru</snm><fnm>P</fnm></au>
    <au><snm>Shamma</snm><fnm>SA</fnm></au>
  </aug>
  <source>The Journal of the Acoustical Society of America</source>
  <publisher>ASA</publisher>
  <pubdate>2005</pubdate>
  <volume>118</volume>
  <issue>2</issue>
  <fpage>887</fpage>
  <lpage>-906</lpage>
</bibl>

<bibl id="B51">
  <title><p>Biomimetic spectro-temporal features for music instrument
  recognition in isolated notes and solo phrases</p></title>
  <aug>
    <au><snm>Patil</snm><fnm>K</fnm></au>
    <au><snm>Elhilali</snm><fnm>M</fnm></au>
  </aug>
  <source>EURASIP J. Audio Speech Music Process.</source>
  <publisher>Nature Publishing Group</publisher>
  <pubdate>2015</pubdate>
  <volume>2015</volume>
  <issue>1</issue>
  <fpage>27</fpage>
</bibl>

<bibl id="B52">
  <title><p>Perceptually salient spectrotemporal modulations for recognition of
  sustained musical instruments</p></title>
  <aug>
    <au><snm>Thoret</snm><fnm>E</fnm></au>
    <au><snm>Depalle</snm><fnm>P</fnm></au>
    <au><snm>McAdams</snm><fnm>S</fnm></au>
  </aug>
  <source>The Journal of the Acoustical Society of America</source>
  <publisher>ASA</publisher>
  <pubdate>2016</pubdate>
  <volume>140</volume>
  <issue>6</issue>
  <fpage>EL478</fpage>
  <lpage>-EL483</lpage>
</bibl>

<bibl id="B53">
  <title><p>Understanding a Deep Machine Listening Model Through Feature
  Inversion</p></title>
  <aug>
    <au><snm>Mishra</snm><fnm>S</fnm></au>
    <au><snm>Sturm</snm><fnm>BL</fnm></au>
    <au><snm>Dixon</snm><fnm>S</fnm></au>
  </aug>
  <source>Proceedings of the International Society on Music Information
  Retrieval (ISMIR) Conference</source>
  <pubdate>2018</pubdate>
  <fpage>755</fpage>
  <lpage>-762</lpage>
</bibl>

<bibl id="B54">
  <title><p>One hundred ways to process time, frequency, rate and scale in the
  central auditory system: a pattern-recognition meta-analysis</p></title>
  <aug>
    <au><snm>Hemery</snm><fnm>E</fnm></au>
    <au><snm>Aucouturier</snm><fnm>JJ</fnm></au>
  </aug>
  <source>Frontiers in computational neuroscience</source>
  <pubdate>2015</pubdate>
  <volume>9</volume>
</bibl>

<bibl id="B55">
  <title><p>Kymatio: Scattering Transforms in Python</p></title>
  <aug>
    <au><snm>Andreux</snm><fnm>M</fnm></au>
    <au><snm>Angles</snm><fnm>T</fnm></au>
    <au><snm>Exarchakis</snm><fnm>G</fnm></au>
    <au><snm>Leonarduzzi</snm><fnm>R</fnm></au>
    <au><snm>Rochette</snm><fnm>G</fnm></au>
    <au><snm>Thiry</snm><fnm>L</fnm></au>
    <au><snm>Zarka</snm><fnm>J</fnm></au>
    <au><snm>Mallat</snm><fnm>S</fnm></au>
    <au><snm>Belilovsky</snm><fnm>E</fnm></au>
    <au><snm>Bruna</snm><fnm>J</fnm></au>
    <au><cnm>others</cnm></au>
  </aug>
  <source>Journal of Machine Learning Research</source>
  <pubdate>2020</pubdate>
  <volume>21</volume>
  <issue>60</issue>
</bibl>

<bibl id="B56">
  <title><p>The Shape of RemiXXXes to Come: Audio texture synthesis with
  time--frequency scattering</p></title>
  <aug>
    <au><snm>Lostanlen</snm><fnm>V</fnm></au>
    <au><snm>Hecker</snm><fnm>F</fnm></au>
  </aug>
  <source>Proceedings of the Digital Audio Effects Conference (DAFX)</source>
  <pubdate>2019</pubdate>
</bibl>

<bibl id="B57">
  <title><p>Understanding deep convolutional networks</p></title>
  <aug>
    <au><snm>Mallat</snm><fnm>S</fnm></au>
  </aug>
  <source>Philosophical Transactions of the Royal Society A: Mathematical,
  Physical and Engineering Sciences</source>
  <publisher>The Royal Society Publishing</publisher>
  <pubdate>2016</pubdate>
  <volume>374</volume>
  <issue>2065</issue>
  <fpage>20150203</fpage>
</bibl>

<bibl id="B58">
  <title><p>Audio content descriptors of timbre</p></title>
  <aug>
    <au><snm>Caetano</snm><fnm>M</fnm></au>
    <au><snm>Saitis</snm><fnm>C</fnm></au>
    <au><snm>Siedenburg</snm><fnm>K</fnm></au>
  </aug>
  <source>Timbre: Acoustics, perception, and cognition</source>
  <publisher>Springer</publisher>
  <pubdate>2019</pubdate>
  <fpage>297</fpage>
  <lpage>-333</lpage>
</bibl>

<bibl id="B59">
  <title><p>A review of automatic drum transcription</p></title>
  <aug>
    <au><snm>Wu</snm><fnm>CW</fnm></au>
    <au><snm>Dittmar</snm><fnm>C</fnm></au>
    <au><snm>Southall</snm><fnm>C</fnm></au>
    <au><snm>Vogl</snm><fnm>R</fnm></au>
    <au><snm>Widmer</snm><fnm>G</fnm></au>
    <au><snm>Hockman</snm><fnm>J</fnm></au>
    <au><snm>Muller</snm><fnm>M</fnm></au>
    <au><snm>Lerch</snm><fnm>A</fnm></au>
  </aug>
  <source>IEEE Transactions on Audio, Speech, and Language Processing</source>
  <publisher>IEEE Press</publisher>
  <pubdate>2018</pubdate>
  <volume>26</volume>
  <issue>9</issue>
  <fpage>1457</fpage>
  <lpage>-1483</lpage>
</bibl>

<bibl id="B60">
  <title><p>Modelling Timbral Hardness</p></title>
  <aug>
    <au><snm>Pearce</snm><fnm>A</fnm></au>
    <au><snm>Brookes</snm><fnm>T</fnm></au>
    <au><snm>Mason</snm><fnm>R</fnm></au>
  </aug>
  <source>Applied Sciences</source>
  <publisher>Multidisciplinary Digital Publishing Institute</publisher>
  <pubdate>2019</pubdate>
  <volume>9</volume>
  <issue>3</issue>
  <fpage>466</fpage>
</bibl>

<bibl id="B61">
  <title><p>Comparison of methods for collecting and modeling dissimilarity
  data: Applications to complex sound stimuli</p></title>
  <aug>
    <au><snm>Giordano</snm><fnm>BL</fnm></au>
    <au><snm>Guastavino</snm><fnm>C</fnm></au>
    <au><snm>Murphy</snm><fnm>E</fnm></au>
    <au><snm>Ogg</snm><fnm>M</fnm></au>
    <au><snm>Smith</snm><fnm>BK</fnm></au>
    <au><snm>McAdams</snm><fnm>S</fnm></au>
  </aug>
  <source>Multivariate behavioral research</source>
  <publisher>Taylor \& Francis</publisher>
  <pubdate>2011</pubdate>
  <volume>46</volume>
  <issue>5</issue>
  <fpage>779</fpage>
  <lpage>-811</lpage>
</bibl>

<bibl id="B62">
  <title><p>Acoustic structure of the five perceptual dimensions of timbre in
  orchestral instrument tones</p></title>
  <aug>
    <au><snm>Elliott</snm><fnm>TM</fnm></au>
    <au><snm>Hamilton</snm><fnm>LS</fnm></au>
    <au><snm>Theunissen</snm><fnm>FE</fnm></au>
  </aug>
  <source>The Journal of the Acoustical Society of America</source>
  <publisher>Acoustical Society of America</publisher>
  <pubdate>2013</pubdate>
  <volume>133</volume>
  <issue>1</issue>
  <fpage>389</fpage>
  <lpage>-404</lpage>
</bibl>

<bibl id="B63">
  <title><p>An efficient heuristic procedure for partitioning
  graphs</p></title>
  <aug>
    <au><snm>Kernighan</snm><fnm>BW</fnm></au>
    <au><snm>Lin</snm><fnm>S</fnm></au>
  </aug>
  <source>The Bell system technical journal</source>
  <publisher>Nokia Bell Labs</publisher>
  <pubdate>1970</pubdate>
  <volume>49</volume>
  <issue>2</issue>
  <fpage>291</fpage>
  <lpage>-307</lpage>
</bibl>

<bibl id="B64">
  <title><p>Scalable parallel data mining for association rules</p></title>
  <aug>
    <au><snm>Han</snm><fnm>EH</fnm></au>
    <au><snm>Karypis</snm><fnm>G</fnm></au>
    <au><snm>Kumar</snm><fnm>V</fnm></au>
  </aug>
  <publisher>ACM</publisher>
  <pubdate>1997</pubdate>
  <volume>26</volume>
  <issue>2</issue>
</bibl>

<bibl id="B65">
  <title><p>Cluster ensembles---a knowledge reuse framework for combining
  multiple partitions</p></title>
  <aug>
    <au><snm>Strehl</snm><fnm>A</fnm></au>
    <au><snm>Ghosh</snm><fnm>J</fnm></au>
  </aug>
  <source>Journal of machine learning research</source>
  <pubdate>2002</pubdate>
  <volume>3</volume>
  <issue>Dec</issue>
  <fpage>583</fpage>
  <lpage>-617</lpage>
</bibl>

<bibl id="B66">
  <title><p>Constant-Q transform toolbox for music processing</p></title>
  <aug>
    <au><snm>Sch{\"o}rkhuber</snm><fnm>C</fnm></au>
    <au><snm>Klapuri</snm><fnm>A</fnm></au>
  </aug>
  <source>Proceedings of the Sound and Music Computing (SMC)
  Conference</source>
  <pubdate>2010</pubdate>
</bibl>

<bibl id="B67">
  <title><p>Learning the helix topology of musical pitch</p></title>
  <aug>
    <au><snm>Lostanlen</snm><fnm>V</fnm></au>
    <au><snm>Sridhar</snm><fnm>S</fnm></au>
    <au><snm>Farnsworth</snm><fnm>A</fnm></au>
    <au><snm>Bello</snm><fnm>JP</fnm></au>
  </aug>
  <source>Proceedings of the International Conference on Acoustics, Speech, and
  Signal Processing (ICASSP)</source>
  <pubdate>2020</pubdate>
</bibl>

<bibl id="B68">
  <title><p>Group invariant scattering</p></title>
  <aug>
    <au><snm>Mallat</snm><fnm>S</fnm></au>
  </aug>
  <source>Communications in Pure and Applied Mathematics</source>
  <publisher>Wiley Online Library</publisher>
  <pubdate>2012</pubdate>
  <volume>65</volume>
  <issue>10</issue>
  <fpage>1331</fpage>
  <lpage>-1398</lpage>
</bibl>

<bibl id="B69">
  <title><p>Unsupervised learning of semantic audio representations</p></title>
  <aug>
    <au><snm>Jansen</snm><fnm>A</fnm></au>
    <au><snm>Plakal</snm><fnm>M</fnm></au>
    <au><snm>Pandya</snm><fnm>R</fnm></au>
    <au><snm>Ellis</snm><fnm>DP</fnm></au>
    <au><snm>Hershey</snm><fnm>S</fnm></au>
    <au><snm>Liu</snm><fnm>J</fnm></au>
    <au><snm>Moore</snm><fnm>RC</fnm></au>
    <au><snm>Saurous</snm><fnm>RA</fnm></au>
  </aug>
  <source>Proceedings of the IEEE International Conference on Acoustics, Speech
  and Signal Processing (ICASSP)</source>
  <pubdate>2018</pubdate>
  <fpage>126</fpage>
  <lpage>-130</lpage>
</bibl>

<bibl id="B70">
  <title><p>{M}etric {L}earning</p></title>
  <aug>
    <au><snm>Bellet</snm><fnm>A</fnm></au>
    <au><snm>Habrard</snm><fnm>A</fnm></au>
    <au><snm>Sebban</snm><fnm>M</fnm></au>
  </aug>
  <publisher>{M}organ \& {C}laypool {P}ublishers</publisher>
  <pubdate>2015</pubdate>
</bibl>

<bibl id="B71">
  <title><p>A rule of thumb: The bandwidth for timbre invariance is one
  octave</p></title>
  <aug>
    <au><snm>Handel</snm><fnm>S</fnm></au>
    <au><snm>Erickson</snm><fnm>ML</fnm></au>
  </aug>
  <source>Music Perception</source>
  <publisher>University of California Press Journals</publisher>
  <pubdate>2001</pubdate>
  <volume>19</volume>
  <issue>1</issue>
  <fpage>121</fpage>
  <lpage>-126</lpage>
</bibl>

<bibl id="B72">
  <title><p>The dependency of timbre on fundamental frequency</p></title>
  <aug>
    <au><snm>Marozeau</snm><fnm>J</fnm></au>
    <au><snm>Cheveign{\'e}</snm><fnm>A</fnm></au>
    <au><snm>McAdams</snm><fnm>S</fnm></au>
    <au><snm>Winsberg</snm><fnm>S</fnm></au>
  </aug>
  <source>The Journal of the Acoustical Society of America</source>
  <publisher>ASA</publisher>
  <pubdate>2003</pubdate>
  <volume>114</volume>
  <issue>5</issue>
  <fpage>2946</fpage>
  <lpage>-2957</lpage>
</bibl>

<bibl id="B73">
  <title><p>Is the bandwidth for timbre invariance only one octave?</p></title>
  <aug>
    <au><snm>Steele</snm><fnm>KM</fnm></au>
    <au><snm>Williams</snm><fnm>AK</fnm></au>
  </aug>
  <source>Music Perception</source>
  <publisher>JSTOR</publisher>
  <pubdate>2006</pubdate>
  <volume>23</volume>
  <issue>3</issue>
  <fpage>215</fpage>
  <lpage>-220</lpage>
</bibl>

<bibl id="B74">
  <title><p>Playing Technique Recognition by Joint Time--Frequency
  Scattering</p></title>
  <aug>
    <au><snm>Wang</snm><fnm>C</fnm></au>
    <au><snm>Lostanlen</snm><fnm>V</fnm></au>
    <au><snm>Benetos</snm><fnm>E</fnm></au>
    <au><snm>Chew</snm><fnm>E</fnm></au>
  </aug>
  <pubdate>2020</pubdate>
</bibl>

<bibl id="B75">
  <title><p>A spectro-temporal modulation index (STMI) for assessment of speech
  intelligibility</p></title>
  <aug>
    <au><snm>Elhilali</snm><fnm>M</fnm></au>
    <au><snm>Chi</snm><fnm>T</fnm></au>
    <au><snm>Shamma</snm><fnm>SA</fnm></au>
  </aug>
  <source>Speech communication</source>
  <publisher>Elsevier</publisher>
  <pubdate>2003</pubdate>
  <volume>41</volume>
  <issue>2-3</issue>
  <fpage>331</fpage>
  <lpage>-348</lpage>
</bibl>

<bibl id="B76">
  <title><p>Detection of speech tokens in noise using adaptive spectrotemporal
  receptive fields</p></title>
  <aug>
    <au><snm>Bellur</snm><fnm>A</fnm></au>
    <au><snm>Elhilali</snm><fnm>M</fnm></au>
  </aug>
  <source>Proceedings of the Annual Conference on Information Sciences and
  Systems (CISS)</source>
  <pubdate>2015</pubdate>
  <fpage>1</fpage>
  <lpage>-6</lpage>
</bibl>

<bibl id="B77">
  <title><p>A multiresolution analysis for detection of abnormal lung
  sounds</p></title>
  <aug>
    <au><snm>Emmanouilidou</snm><fnm>D</fnm></au>
    <au><snm>Patil</snm><fnm>K</fnm></au>
    <au><snm>West</snm><fnm>J</fnm></au>
    <au><snm>Elhilali</snm><fnm>M</fnm></au>
  </aug>
  <source>Proceedings of the International Conference of the IEEE Engineering
  in Medicine and Biology Society (EMBS)</source>
  <pubdate>2012</pubdate>
  <fpage>3139</fpage>
  <lpage>-3142</lpage>
</bibl>

<bibl id="B78">
  <title><p>A dictionary of economics</p></title>
  <aug>
    <au><snm>Black</snm><fnm>J</fnm></au>
    <au><snm>Hashimzade</snm><fnm>N</fnm></au>
    <au><snm>Myles</snm><fnm>G</fnm></au>
  </aug>
  <publisher>Oxford university press</publisher>
  <pubdate>2012</pubdate>
</bibl>

<bibl id="B79">
  <title><p>OrchideaSOL: A Dataset of Extended Instrumental Techniques for
  Computer-aided Orchestration</p></title>
  <aug>
    <au><snm>Cella</snm><fnm>CE</fnm></au>
    <au><snm>Ghisi</snm><fnm>D</fnm></au>
    <au><snm>Lostanlen</snm><fnm>V</fnm></au>
    <au><snm>L{\'e}vy</snm><fnm>F</fnm></au>
    <au><snm>Fineberg</snm><fnm>J</fnm></au>
    <au><snm>Maresz</snm><fnm>Y</fnm></au>
  </aug>
  <source>Proceedings of the International Computer Music Conference
  (ICMC)</source>
  <pubdate>2020</pubdate>
</bibl>

<bibl id="B80">
  <title><p>Modeling the onset advantage in musical instrument
  recognition</p></title>
  <aug>
    <au><snm>Siedenburg</snm><fnm>K</fnm></au>
    <au><snm>Sch{\"a}dler</snm><fnm>MR</fnm></au>
    <au><snm>H{\"u}lsmeier</snm><fnm>D</fnm></au>
  </aug>
  <source>The journal of the Acoustical Society of America</source>
  <publisher>Acoustical Society of America</publisher>
  <pubdate>2019</pubdate>
  <volume>146</volume>
  <issue>6</issue>
  <fpage>EL523</fpage>
  <lpage>-EL529</lpage>
</bibl>

<bibl id="B81">
  <title><p>On Time-frequency Scattering and Computer Music</p></title>
  <aug>
    <au><snm>Lostanlen</snm><fnm>V</fnm></au>
  </aug>
  <source>Florian Hecker: Halluzination, Perspektive, Synthese</source>
  <publisher>Berlin: Sternberg Press</publisher>
  <editor>Schafhausen, Nicolaus and M\"{u}ller, Vanessa Joan</editor>
  <pubdate>2019</pubdate>
</bibl>

<bibl id="B82">
  <title><p>Representing environmental sounds using the separable scattering
  transform</p></title>
  <aug>
    <au><snm>Baug{\'e}</snm><fnm>C</fnm></au>
    <au><snm>Lagrange</snm><fnm>M</fnm></au>
    <au><snm>And{\'e}n</snm><fnm>J</fnm></au>
    <au><snm>Mallat</snm><fnm>S</fnm></au>
  </aug>
  <source>Proceedings of the IEEE International Conference on Acoustics, Speech
  and Signal Processing</source>
  <pubdate>2013</pubdate>
  <fpage>8667</fpage>
  <lpage>-8671</lpage>
</bibl>

<bibl id="B83">
  <title><p>Adaptive Time--Frequency Scattering for Periodic Modulation
  Recognition in Music Signals</p></title>
  <aug>
    <au><snm>Wang</snm><fnm>C</fnm></au>
    <au><snm>Benetos</snm><fnm>E</fnm></au>
    <au><snm>Lostanlen</snm><fnm>V</fnm></au>
    <au><snm>Chew</snm><fnm>E</fnm></au>
  </aug>
  <pubdate>2019</pubdate>
</bibl>

<bibl id="B84">
  <title><p>Separable spectro-temporal Gabor filter bank features: Reducing the
  complexity of robust features for automatic speech recognition</p></title>
  <aug>
    <au><snm>Sch{\"a}dler</snm><fnm>MR</fnm></au>
    <au><snm>Kollmeier</snm><fnm>B</fnm></au>
  </aug>
  <source>The Journal of the Acoustical Society of America</source>
  <publisher>Acoustical Society of America</publisher>
  <pubdate>2015</pubdate>
  <volume>137</volume>
  <issue>4</issue>
  <fpage>2047</fpage>
  <lpage>-2059</lpage>
</bibl>

<bibl id="B85">
  <title><p>Training and testing low-degree polynomial data mappings via linear
  SVM</p></title>
  <aug>
    <au><snm>Chang</snm><fnm>YW</fnm></au>
    <au><snm>Hsieh</snm><fnm>CJ</fnm></au>
    <au><snm>Chang</snm><fnm>KW</fnm></au>
    <au><snm>Ringgaard</snm><fnm>M</fnm></au>
    <au><snm>Lin</snm><fnm>CJ</fnm></au>
  </aug>
  <source>Journal of Machine Learning Research</source>
  <pubdate>2010</pubdate>
  <volume>11</volume>
  <issue>Apr</issue>
  <fpage>1471</fpage>
  <lpage>-1490</lpage>
</bibl>

</refgrp>
} 

\begin{table}
\caption{Full list of audio stimuli (1/2). In each instrument, a blank space in the rightmost column denotes the ordinary playing technique (\emph{ordinario}).}
\label{tab:list1}
      \begin{tabular}{llll}
          name & instrument  & mute   &  playing technique \\ \hline \hline
        \texttt{ASax-key-cl-C4-p} & Alto saxophone & & key click \\
        \texttt{ASax-ord-C4-mf} & Alto saxophone &  &  \\
        \texttt{ASax-ord-hi-reg-C6-mf} & Alto saxophone &   & \\
        \texttt{ASax-slap-C4-mf} & Alto saxophone & & slap tonguing \\
        \texttt{ASax-slap-unp-C4-p} & Alto saxophone & & unpitched slap tonguing \\ \hline
        \texttt{BbCl-key-cl-C4-pp} & Soprano clarinet & & key click \\
        \texttt{BbCl-ord-hi-reg-A6-ff} & Soprano clarinet & & \\ \hline
        \texttt{BBTb-explo-slap-C\#1-mf} & Bass tuba &  & explosive slap \\
		\texttt{BBTb-pdl-tone-F1-mf} & Bass tuba & & pedal tone \\
		\texttt{BBTb-slap-F1-mf} & Bass tuba & & slap tonguing \\
		\texttt{BBTb-slap-unp-mf-1} & Bass tuba & & unpitched slap tonguing \\ \hline
		\texttt{Bn-key-cl-C3-mf} & Bassoon & & key click \\
		\texttt{Bn-ord-C4-mf} & Bassoon & & \\ \hline
		\texttt{Cb-pizz-bartok-C4-ff-1c} & Contrabass & & pizzicato \`a la Bart\'ok (snap) \\
		\texttt{Cb-pizz-lv-C4-mf-1c} & Contrabass & & pizzicato \& \emph{laissez vibrer} \\
        \texttt{Cb-pizz-sec-C4-mf-1c} & Contrabass & & pizzicato secco \\
        \texttt{Cb-pont-C4-mf-1c} & Contrabass & & sul ponticello \\ \hline
        \texttt{Fl-key-cl-C4-f} & Flute & & key click \\
        \texttt{Fl-ord-C4-mf} & Flute & & \\
        \texttt{Fl-tng-ram-C4-mf} & Flute & & tongue ram \\ \hline
        \texttt{Gtr-ord-C4-mf-2c} & Spanish guitar & & \\
        \texttt{Gtr-ord-hi-reg-E5-mf-3c} & Spanish guitar & & \\
        \texttt{Gtr-pizz-bartok-C4-ff-2c} & Spanish guitar & & pizzicato Bart\'ok\\
        \texttt{Gtr-pizz-C4-mf-2c} & Spanish guitar &  & pizzicato \\ \hline
        \texttt{Hn-ord-C4-mf} & French horn & & \\
        \texttt{Hn-slap-C4-mf} & French horn & & slap tonguing \\ \hline
        \texttt{Hp-harm-fngr-C4-f} & Harp & & harmonic fingering \\
        \texttt{Hp-ord-C4-m4} & Harp & & \\
        \texttt{Hp-pizz-bartok-C4-mf} & Harp & & pizzicato Bart\'ok \\
        \texttt{Hp-xyl-C4-p} & Harp & & xylophonic \\ \hline
        \texttt{Ob-blow-no-reed-C4} & Oboe & & blow without reed \\
        \texttt{Ob-key-cl-C4-pp} & Oboe & & key click \\
        \texttt{Ob-ord-C4-mf} & Oboe & & \\ \hline
        \texttt{TpC-ord-C4-mf} & Trumpet in C & & \\
        \texttt{TpC-pdl-tone-F3-mf} & Trumpet in C & & pedal tone \\
        \texttt{TpC-slap-C4-p} & Trumpet in C & & slap tonguing \\
        \texttt{TpC+C-ord-C4-mf} & Trumpet in C & cup & \\
        \texttt{TpC+H-ord-C4-mf} & Trumpet in C & harmon & \\
        \texttt{TpC+S-ord-C4-mf} & Trumpet in C & straight & \\
        \texttt{TpC+W-ord-closed-C4-mf} & Trumpet in C & wah (closed) & \\
        \texttt{TpC+W-ord-open-C4-mf} & Trumpet in C & wah (open) & \\ \hline
        \texttt{TTbn-ord-C4-mf} & Tenor trombone & &  \\
        \texttt{TTbn+C-ord-C4-mf} & Tenor trombone & cup & \\
        \texttt{TTbn+H-ord-C4-mf} & Tenor trombone & harmon & \\
        \texttt{TTbn+S-ord-C4-mf} & Tenor trombone & straight & \\
        \texttt{TTbn+W-ord-closed-C4-mf} & Tenor trombone & wah (closed) & \\
        \texttt{TTbn+W-ord-open-C4-mf} & Tenor trombone & wah (open) & \\ \hline
        \texttt{Va-art-harm-C5-mf-4c} & Viola & & artificial harmonic \\
        \texttt{Va-legno-batt-C4-mf-3c} & Viola & & col legno battuto \\
        \texttt{Va-ord-C4-mf-3c} & Viola & & \\
        \texttt{Va-pizz-bartok-C4-ff-3c} & Viola & & pizzicato Bart\'ok \\
        \texttt{Va-pizz-lv-C4-mf-3c} & Viola & & pizzicato \& \emph{laissez vibrer} \\
        \texttt{Va-pizz-sec-C4-mf-3c} & Viola & & pizzicato secco \\
        \texttt{Va-pont-C4-mf-3c} & Viola & & sul ponticello \\
        \texttt{Va+S-ord-C3-mf-3c} & Viola & sordina & \\
        \texttt{Va+SP-ord-D4-mf-2c} & Viola & piombo & \\ \hline
        \texttt{Vc-art-harm-C4-mf} & Cello & & artificial harmonic \\
        \texttt{Vc-legno-batt-C4-mf-1c} & Cello & & col legno battuto \\
        \texttt{Vc-legno-tratto-C4-mf-1c} & Cello & & col legno tratto \\
        \texttt{Vc-nonvib-C4-mf-1c} & Cello & & nonvibrato \\
        \texttt{Vc-ord-C4-mf-1c} & Cello & & \\
        \texttt{Vc-pizz-bartok-C4-ff-1c} & Cello & & pizzicato Bart\'ok \\
        \texttt{Vc-pizz-lv-C4-mf-1c} & Cello & & pizzicato \& \emph{laissez vibrer} \\
        \texttt{Vc-pizz-sec-C4-mf-1c} & Cello & & pizzicato secco \\
        \texttt{Vc-pont-C4-mf-2c} & Cello & & sul ponticello \\
        \texttt{Vc-tasto-C4-mf-1c} & Cello & & sul tasto \\
        \texttt{Vc+S-ord-C4-mf-1c} & Cello & sordina & \\
        \texttt{Vc+SP-ord-C4-mf-1c} & Cello & piombo & \\ \hline
    \end{tabular}
\end{table}

\begin{table*}
\caption{Full list of audio stimuli (2/2). In each instrument, a blank space in the rightmost column denotes the ordinary playing technique (\emph{ordinario}).}
\label{tab:list2}
\begin{tabular}{llll}
          name & instrument  & mute   &  playing technique \\ \hline \hline
          \texttt{Vn-art-harm-G5-mf-4c} & Violin & & artificial harmonic \\
          \texttt{Vn-legno-batt-C4-mf-4c} & Violin & & col legno battuto \\
          \texttt{Vn-nonvib-C4-mf-4c} & Violin & & nonvibrato \\
          \texttt{Vn-ord-C4-mf-4c} & Violin & & \\
          \texttt{Vn-pizz-bartok-C4-ff-4c} & Violin & & pizzicato Bart\'ok \\
          \texttt{Vn-pizz-lv-C4-mf-4c} & Violin & pizzicato \& \emph{laissez vibrer} \\
          \texttt{Vn-pizz-sec-C4-mf-4c} & Violin & pizzicato secco \\
          \texttt{Vn-pont-C4-mf-4c} & Violin & & sul ponticello \\
          \texttt{Vn+S-ord-C4-mf-4c} & Violin & sordina & \\
          \texttt{Vn+SP-ord-C4-mf-4c} & Violin & piombo & \\ \hline
\end{tabular}
\end{table*}

\section*{Figures}

\begin{figure}[h!]
\includegraphics[width=0.9\textwidth]{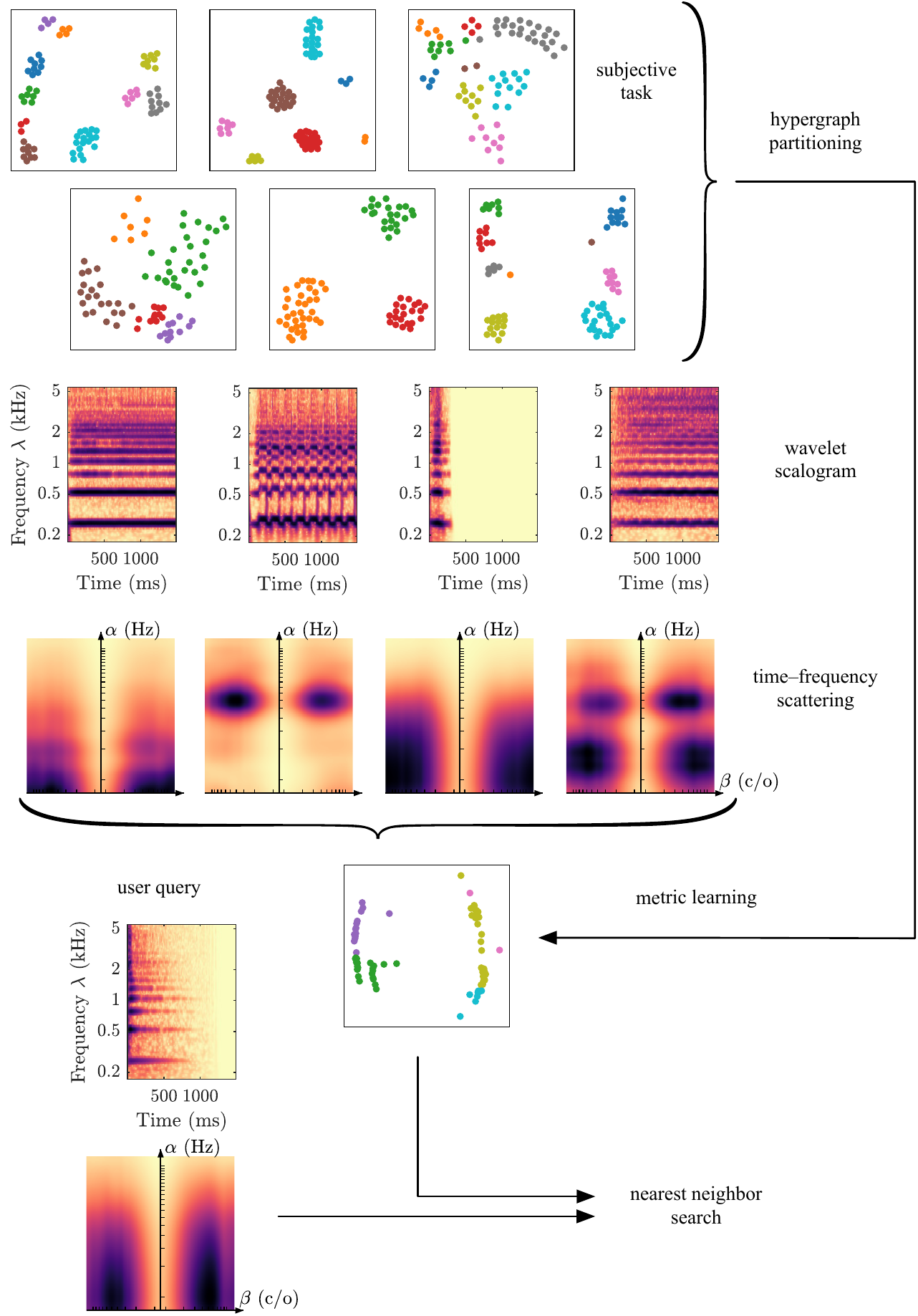}
\caption{\csentence{Overview of the proposed approach.} See \lnameref{sec:intro} for details.}
\label{fig:pipeline}
\end{figure}

\begin{figure}
\center
\begin{tabular}{cc}
  (a) & (b) \\
  \includegraphics[
  	width = 0.47\textwidth, trim = 10mm 0 5mm 0, clip]
	  {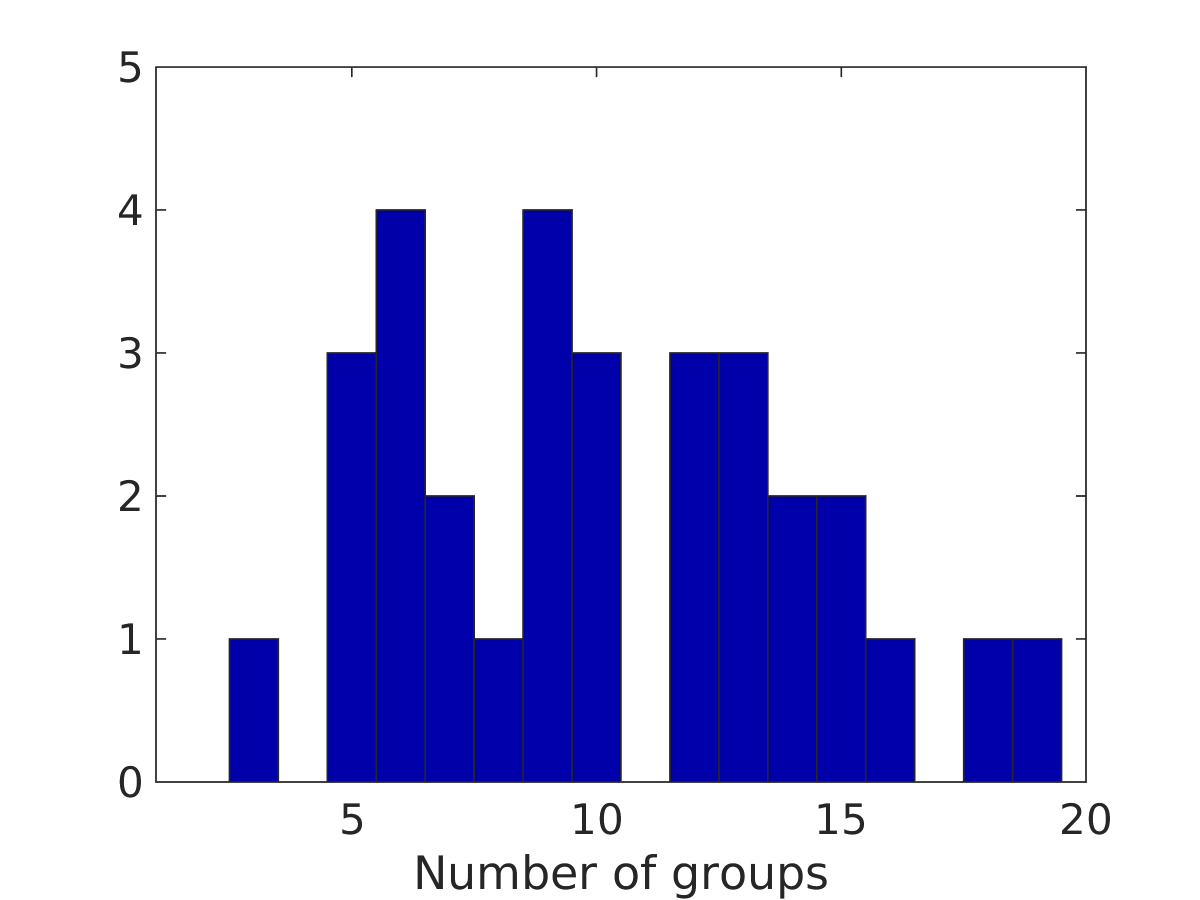} &
  \includegraphics[
    width = 0.47\textwidth, trim = 10mm 0 5mm 0, clip]
      {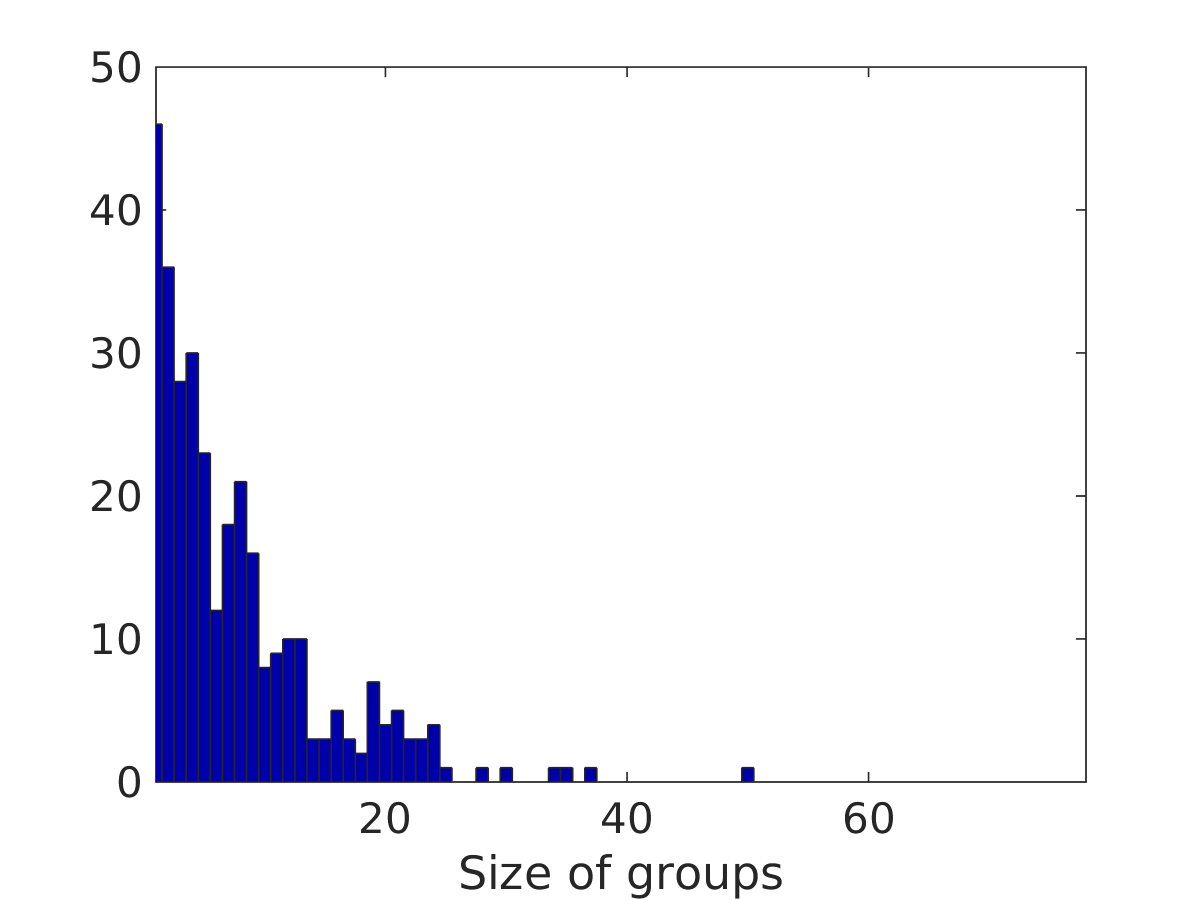}
\end{tabular}
\caption{\csentence{Inter-subject variability}.
Histogram of the number of clusters (a) and the size of the clusters (b) defined by the $31$ subjects. 
See \lnameref{sec:data-collection} for details.}
\label{fig:cluster-histograms}
\end{figure}

\begin{figure}
\includegraphics[height=4cm]{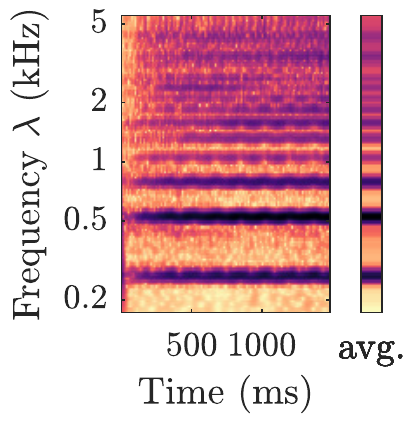}
\hspace{8mm}
\includegraphics[trim=35 0 0 0, clip, height=4cm]{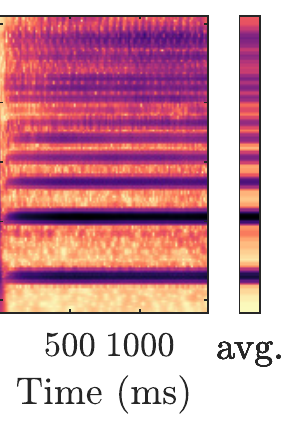}
\hspace{12mm}
\includegraphics[trim=35 0 0 0, clip, height=4cm]{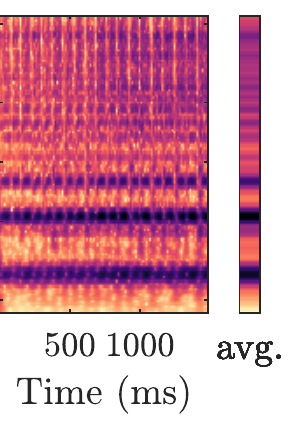}

\hspace{2mm}
\includegraphics[height=28mm]{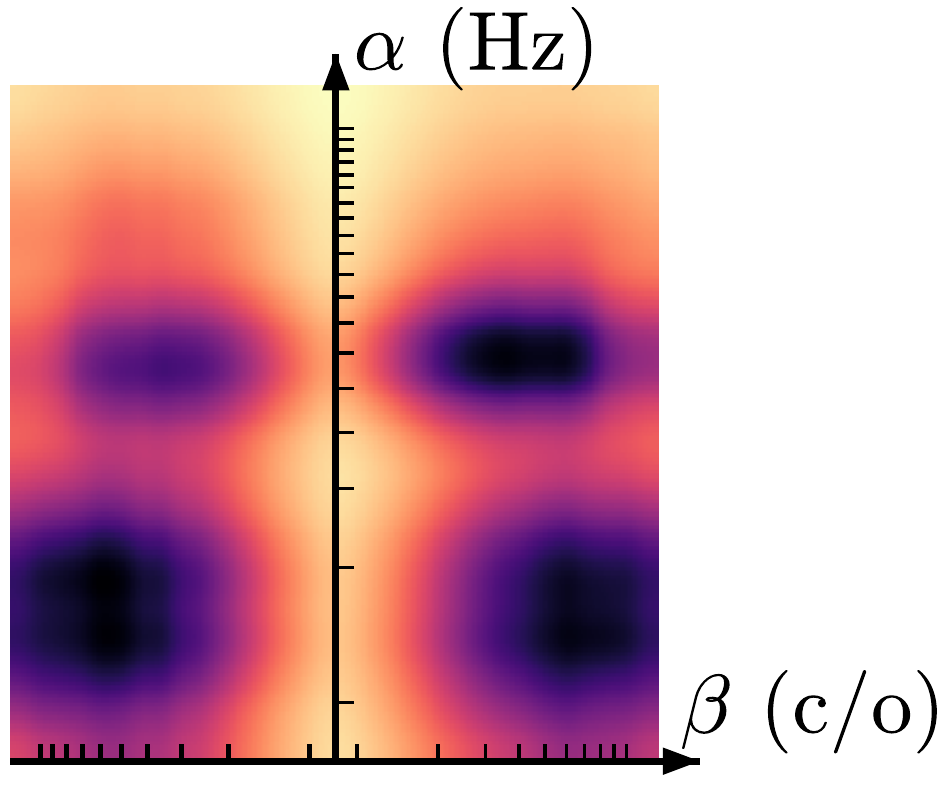}
\hspace{5mm}
\includegraphics[height=28mm]{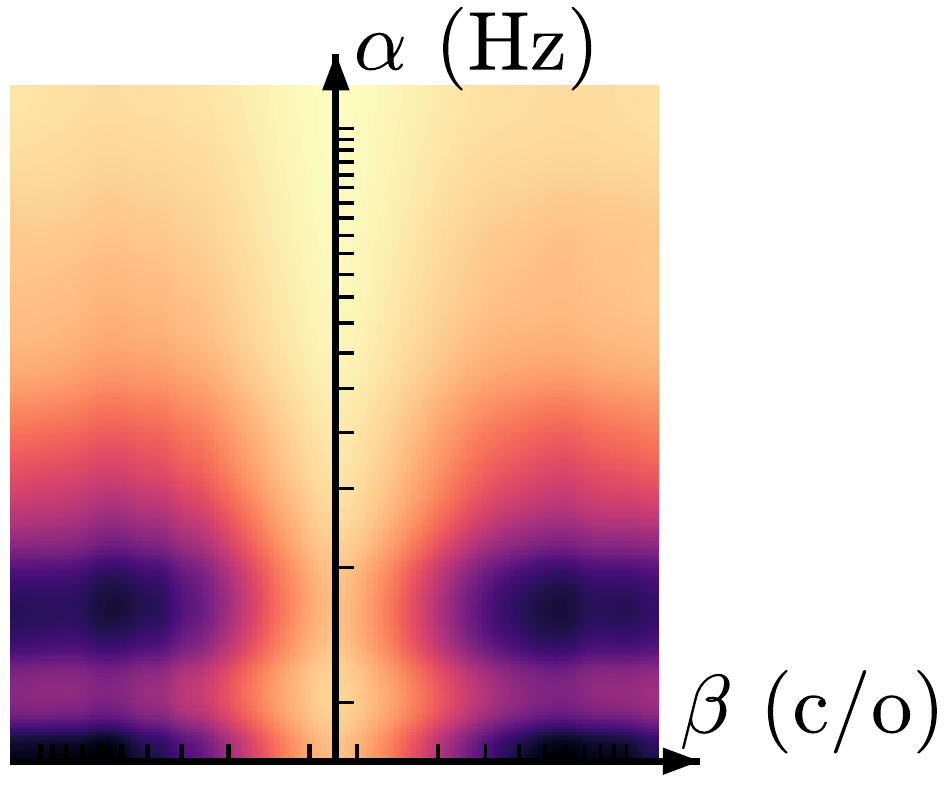}
\hspace{5mm}
\includegraphics[trim=0 0 77 0, clip, height=28mm]{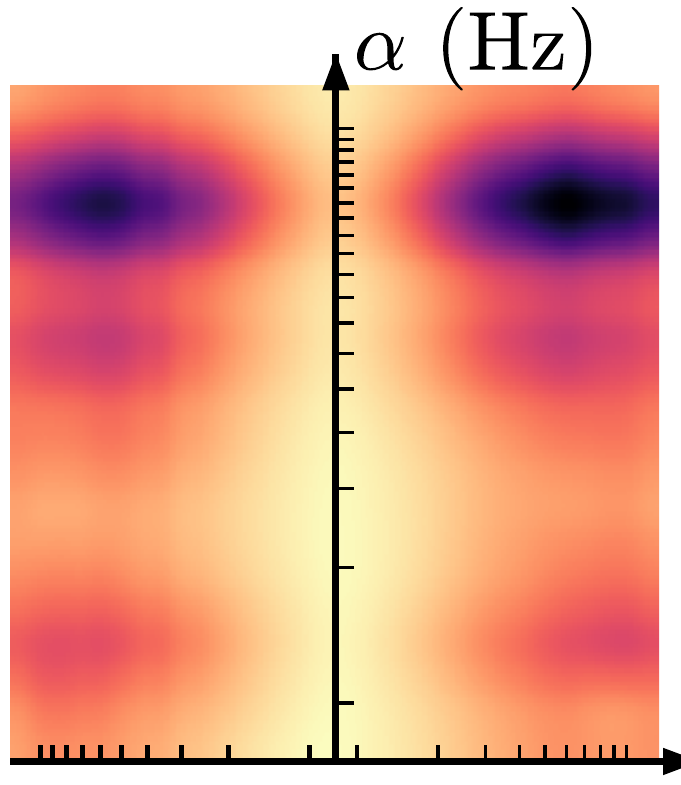}

\csentence{(a) ordinario (with vibrato)}\hspace{18mm}
\csentence{(b) nonvibrato}\hspace{24mm}
\csentence{(c) tremolo}\hspace{8mm}
\vspace{1cm}

\includegraphics[height=4cm]{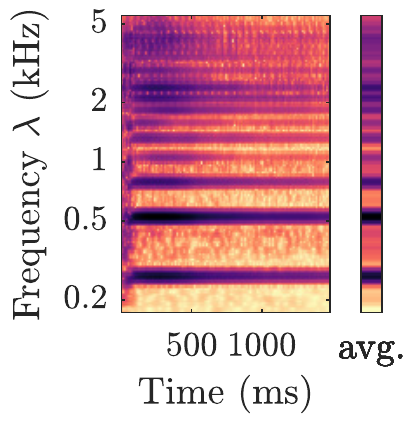}
\hspace{8mm}
\includegraphics[trim=35 0 0 0, clip, height=4cm]{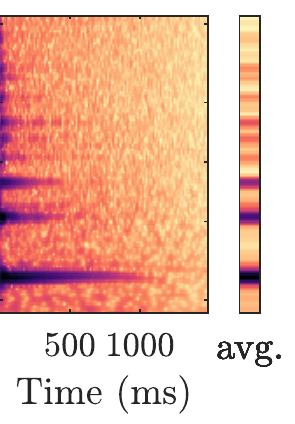}
\hspace{12mm}
\includegraphics[trim=35 0 0 0, clip, height=4cm]{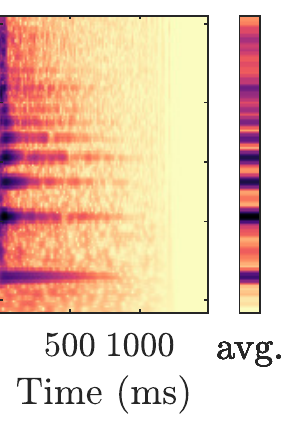}

\hspace{2mm}
\includegraphics[height=28mm]{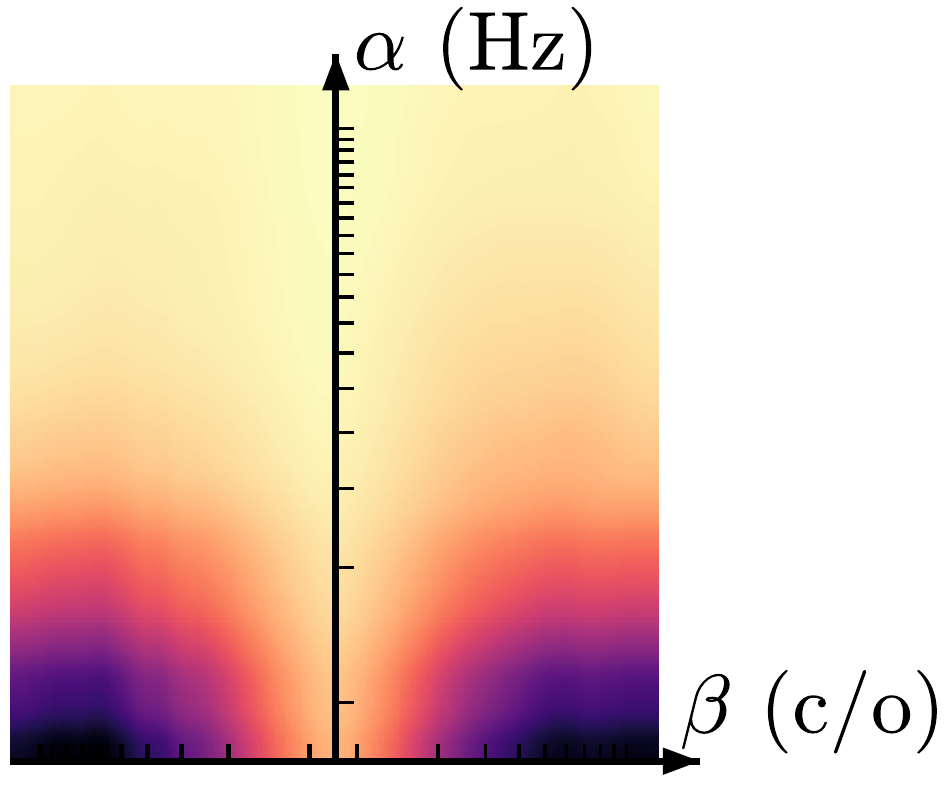}
\hspace{5mm}
\includegraphics[height=28mm]{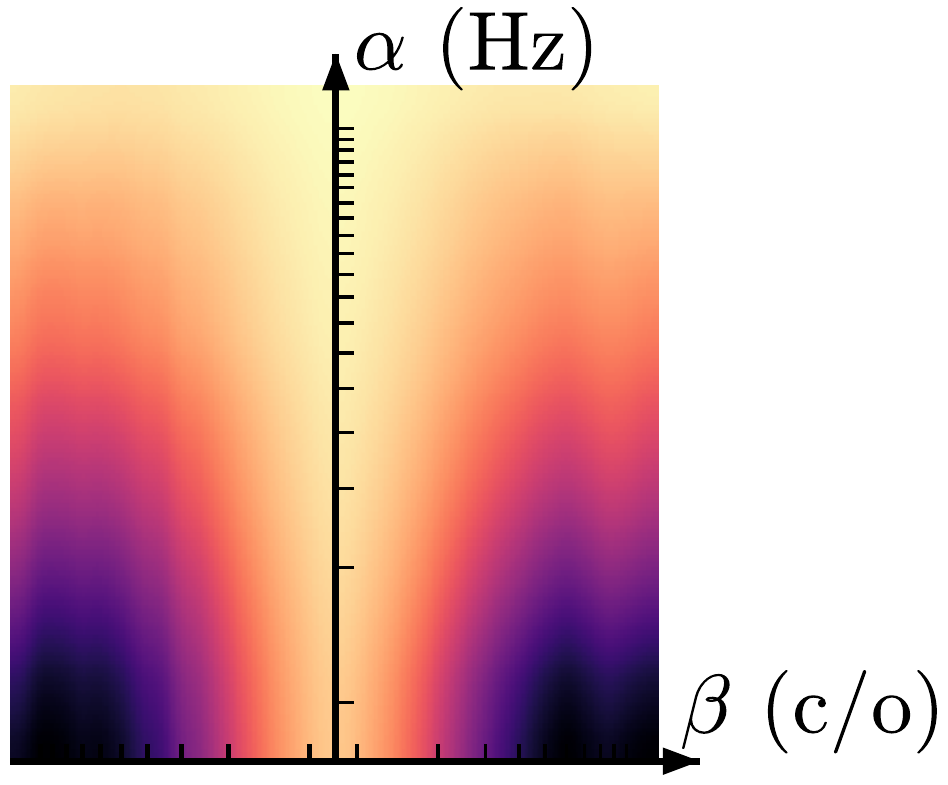}
\hspace{5mm}
\includegraphics[trim=0 0 77 0, clip, height=28mm]{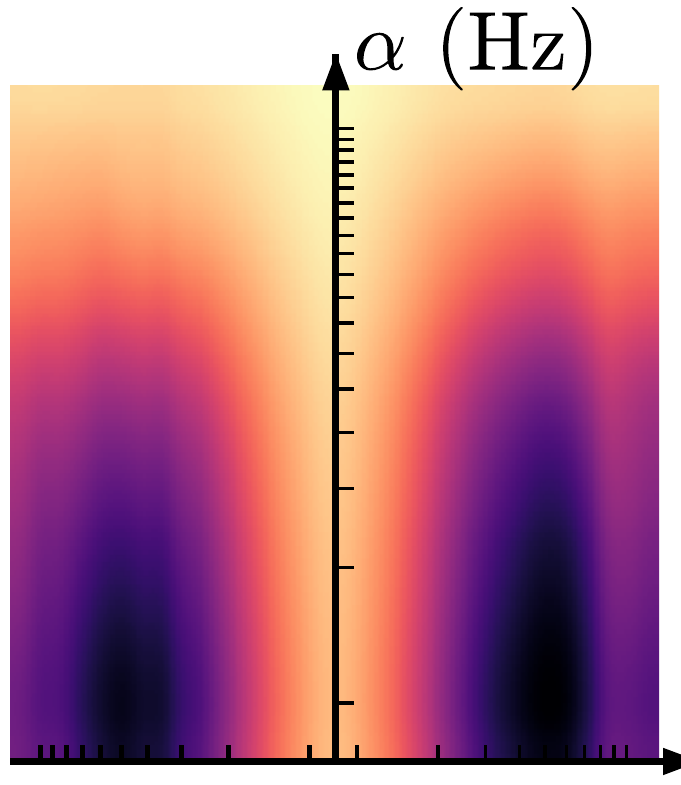}

\hspace{2mm}
\csentence{(d) sforzando}\hspace{25mm}
\csentence{(e) pizzicato}\hspace{25mm}
\csentence{(f) staccato}\hspace{1mm}

\caption{\csentence{Six playing techniques of the violin.}
Subfigures:
(a) ordinario (with vibrato), (b) nonvibrato, (c) tremolo, (d) sforzando, (e) pizzicato (\emph{laissez vibrer}, i.e., let ring), and (f) staccato.
In each subfigure, the top image shows the wavelet scalogram as a function of time $t$ (in seconds) and frequency $\lambda$ (in Hertz).
Conversely, the bottom image shows the average time--frequency scattering coefficients, as a function of temporal modulation rate $\alpha$ (in Hertz) and frequential modulation scale $\beta$ (in cycles per octave).
Darker shades denote greater values of acoustic energy.
See \lnameref{sec:methods} section for details.}
\label{fig:violin-scattering}
\end{figure}

\begin{figure}
\includegraphics[height=4cm]{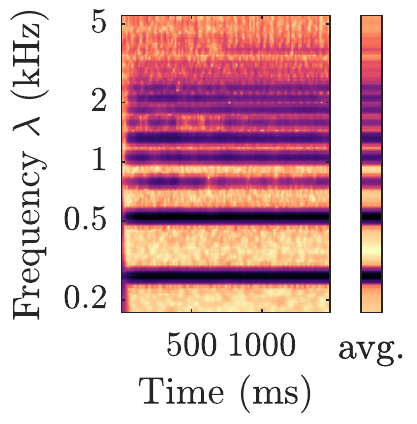}
\hspace{8mm}
\includegraphics[trim=35 0 0 0, clip, height=4cm]{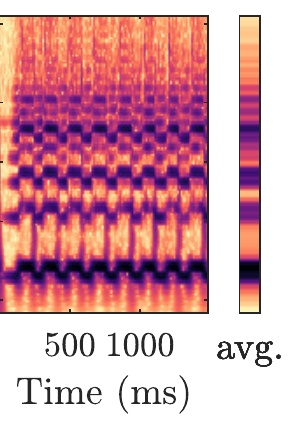}
\hspace{12mm}
\includegraphics[trim=35 0 0 0, clip, height=4cm]{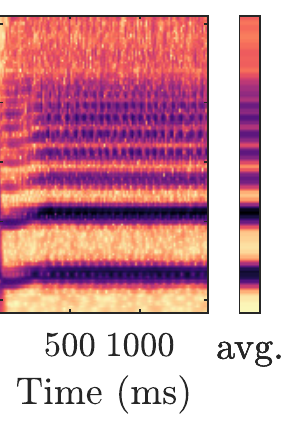}

\hspace{2mm}
\includegraphics[height=28mm]{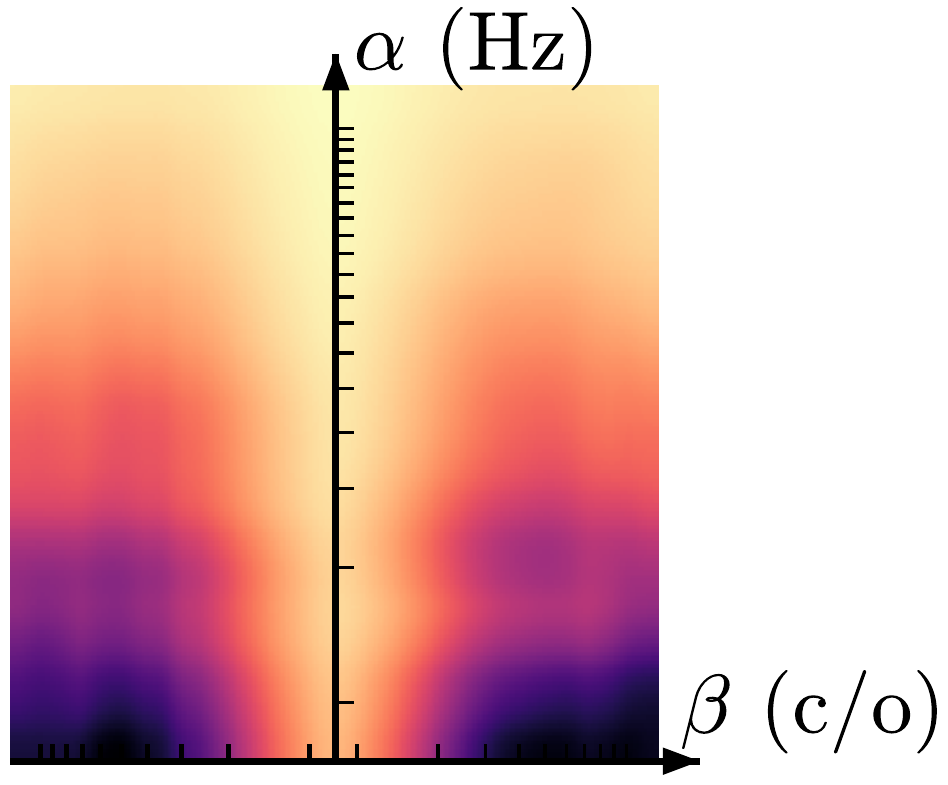}
\hspace{5mm}
\includegraphics[height=28mm]{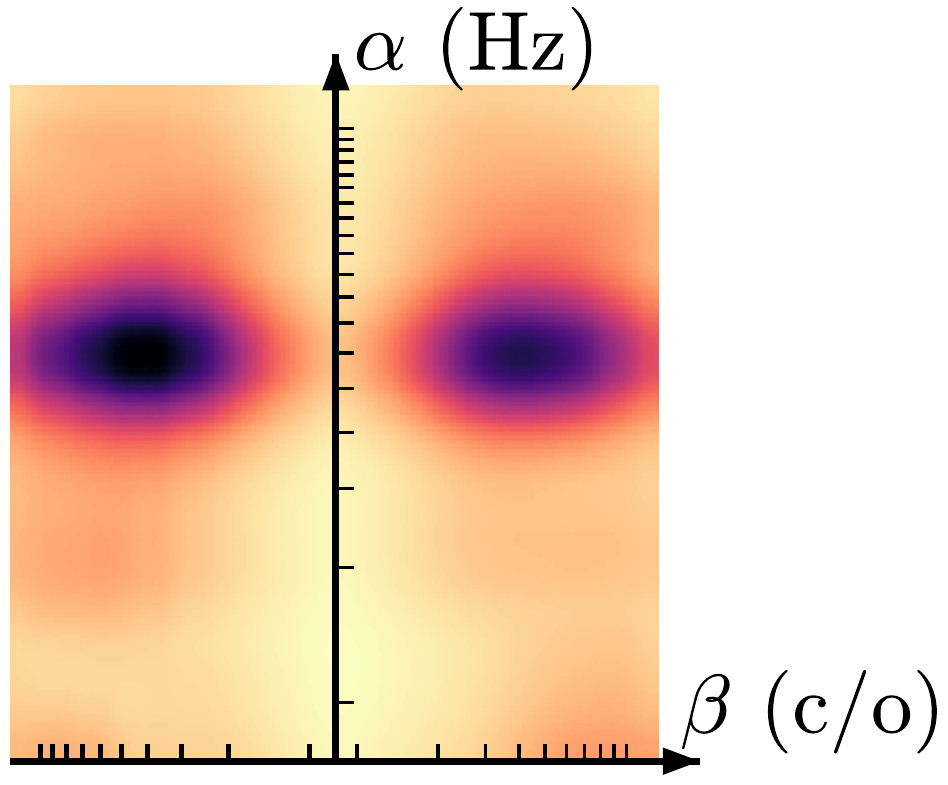}
\hspace{5mm}
\includegraphics[trim=0 0 77 0, clip, height=28mm]{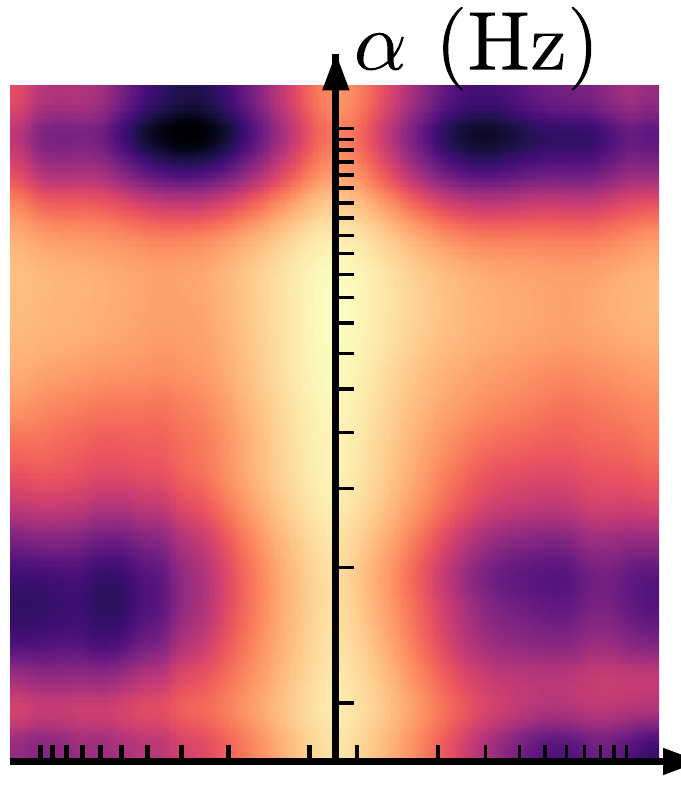}

\hspace{12mm}
\csentence{(a) ordinario}\hspace{22mm}
\csentence{(b) trill $\mathsf{C_4} - \mathsf{D_4}$}\hspace{15mm}
\csentence{(c) play $\mathsf{C\musSharp_4}$ while singing $\mathsf{C_4}$}\vspace{1cm}

\includegraphics[height=4cm]{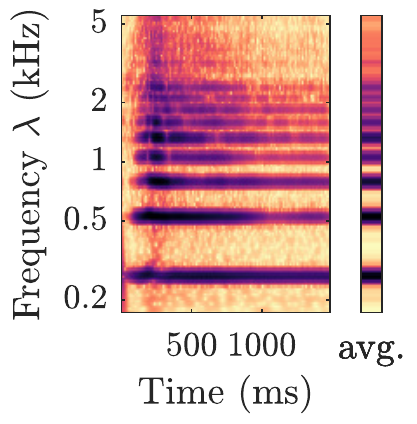}
\hspace{8mm}
\includegraphics[trim=35 0 0 0, clip, height=4cm]{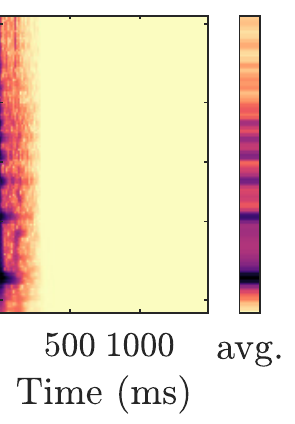}
\hspace{12mm}
\includegraphics[trim=35 0 0 0, clip, height=4cm]{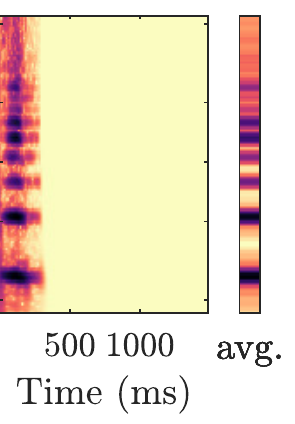}

\hspace{2mm}
\includegraphics[height=28mm]{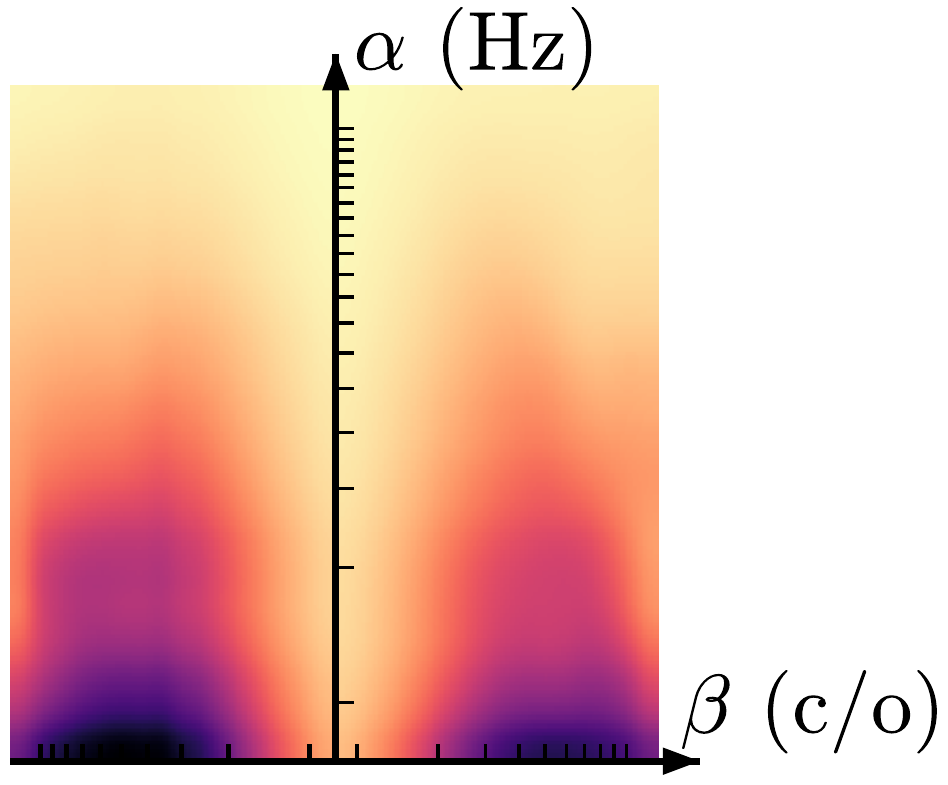}
\hspace{5mm}
\includegraphics[height=28mm]{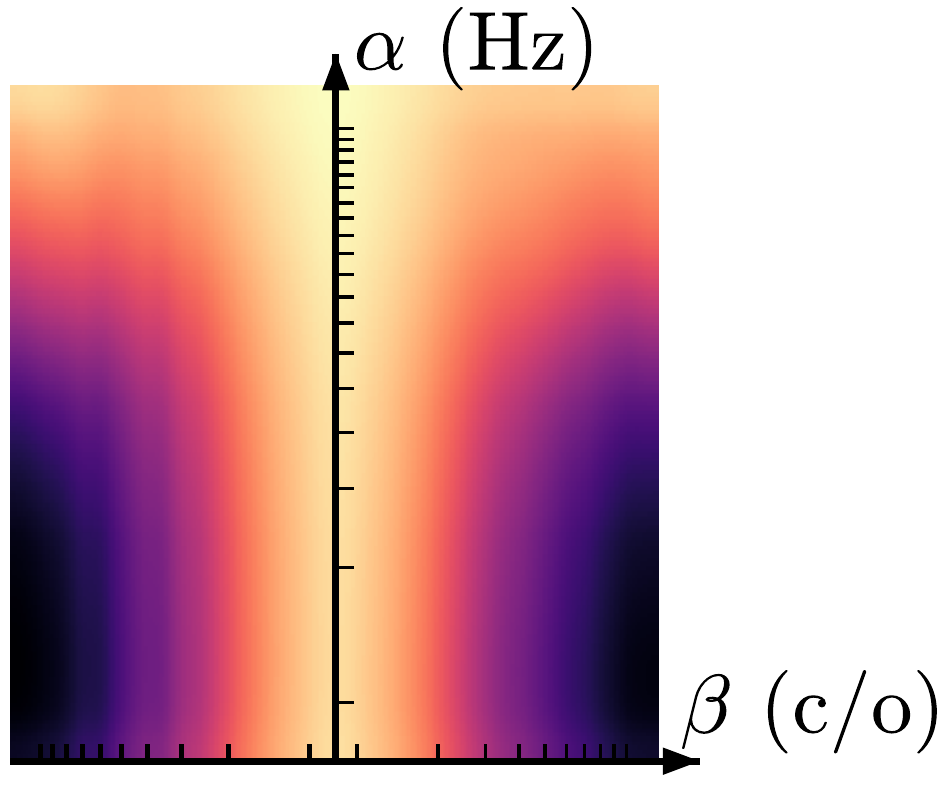}
\hspace{5mm}
\includegraphics[trim=0 0 77 0, clip, height=28mm]{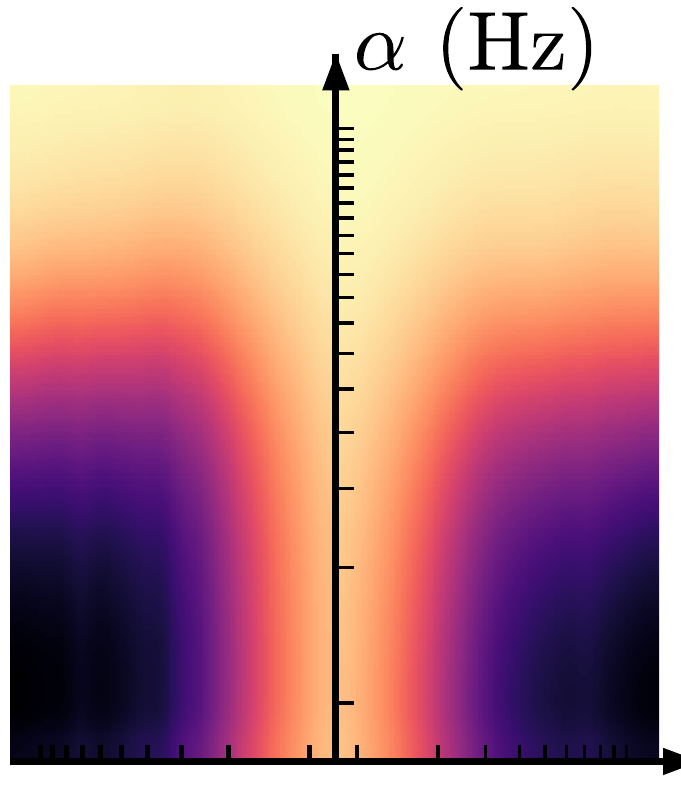}

\hspace{2mm}
\csentence{(d) sforzando}\hspace{25mm}
\csentence{(e) key click}\hspace{25mm}
\csentence{(f) staccato}\hspace{1mm}

\caption{\csentence{Six playing techniques of the flute.}
Subfigures:
(a) ordinario, (b) trill ($\mathsf{C_4} - \mathsf{D_4}$), (c) interference (play $\mathsf{C\musSharp_4}$ while singing $\mathsf{C_4}$), (d) sforzando, (e) key click, (f) staccato.
In each subfigure, the top image shows the wavelet scalogram as a function of time $t$ (in seconds) and frequency $\lambda$ (in Hertz).
Conversely, the bottom image shows the average time--frequency scattering coefficients, as a function of temporal modulation rate $\alpha$ (in Hertz) and frequential modulation scale $\beta$ (in cycles per octave).
Darker shades denote greater values of acoustic energy.
See \lnameref{sec:methods} section for details.}
\label{fig:flute-scattering}
\end{figure}

\begin{figure}[h!]
\includegraphics[width=0.9\textwidth]{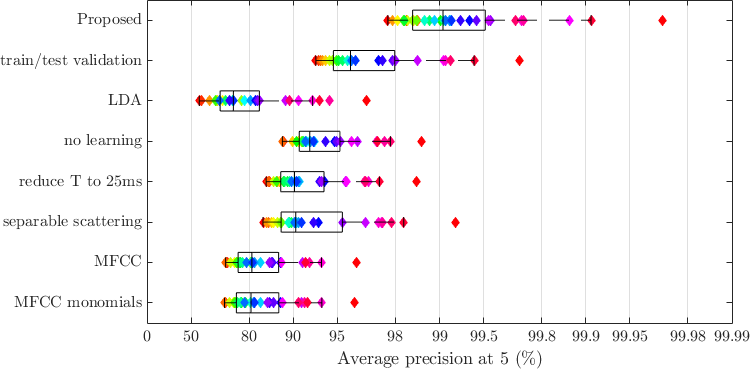}
\caption{\csentence{Impact of different processing architecture or protocol designs.} For each condition, the central mark indicates the median, and the bottom and top edges of the box indicate the 25th and 75th percentiles, respectively. The performance achieved for each reference clustering is depicted by a lozenge whose color is chosen arbitrarily but consistently across conditions.
See \lnameref{sec:results} section for details.}
\label{fig:ablation}
\end{figure}






\end{backmatter}

\end{document}